\documentclass{article}

\usepackage{arxiv}

\usepackage[utf8]{inputenc} 
\usepackage[T1]{fontenc}    
\usepackage{natbib}
\usepackage{hyperref}       
\usepackage{url}            
\usepackage{booktabs}       
\usepackage{amsfonts}       
\usepackage{nicefrac}       
\usepackage{microtype}      
\usepackage{lipsum}
\usepackage{graphicx}
\usepackage{multirow}
\usepackage{subfigure}
\usepackage{subcaption}
\usepackage{graphicx}
\graphicspath{ {./images/} }

\title{Talk, Listen, Connect: How Humans and AI Evaluate Empathy in Responses to Emotionally Charged Narratives}

\author{
 Mahnaz Roshanaei \\
  Department of Communication\\
  Stanford University\\
   \And
 Rezvaneh Rezapour \\
  School of Coumputing and Informatic\\
  Drexel University\\
  \And
 Magy Seif El-Nasr \\
  Department of Computational Media\\
  University of California, Santa Cruz\\
}

\begin{document}
\maketitle
\begin{abstract}
Social interactions promote well-being, yet barriers like geographic distance, time limitations, and mental health conditions can limit face-to-face interactions. Emotionally responsive AI systems, such as chatbots, offer new opportunities for social and emotional support, but raise critical questions about how empathy is perceived and experienced in human-AI interactions. This study examines how empathy is evaluated in AI-generated versus human responses. Using personal narratives, we explored how persona attributes (e.g., gender, empathic traits, shared experiences) and story qualities affect empathy ratings. We compared responses from standard and fine-tuned AI models with human judgments. Results show that while humans are highly sensitive to emotional vividness and shared experience, AI-responses are less influenced by these cues, often lack nuance in empathic expression. These findings highlight challenges in designing emotionally intelligent systems that respond meaningfully across diverse users and contexts, and informs the design of ethically aware tools to support social connection and well-being.  
\end{abstract}


\section{Introduction}
Research consistently demonstrates that a rich social life with support networks and engaging in high-quality social interactions, in particular, face-to-face interactions, are associated with a variety of benefits to people's well-being \cite{cohen1985stress, sandstrom2014social,siedlecki2014relationship,webster2021association}. 
Engaging in social interactions that are meaningful and substantive, such as those involving self-disclosure or emotional depth, have been linked with greater happiness, life satisfaction, and social connectedness \cite{sun2020well, mehl2010eavesdropping, milek2018eavesdropping, roshanaei2024meaningful}. Face-to-face and hybrid interactions consistently outperform digital-only exchanges in promoting positive affect \cite{kroencke2023well}.
Despite all the benefits of face-to-face interactions, barriers such as geographical distance, time constraints, health challenges, social anxiety and loneliness often limit these engagements \cite{matthews2019lonely, segrin1994negative, kashdan2010darker}.
In response to these barriers, AI-driven chatbots have emerged as supplementary tools for facilitating social interaction and offering non-judgmental and accessible support for social, emotional, and relational needs \cite{zhou2020design}, though not as replacements for human contact \cite{li2023feasibility, lee2021social}. 

As chatbots become more fluent and emotionally responsive \cite{paiva2017empathy, yalccin2019evaluating}, questions are raised about their psychological impact, especially during emotionally complex exchanges. Some worry these tools may alter our understanding of intimacy or authenticity in social interactions,potentially reshaping the concept of conceitedness in a digital era.  
Studies on emotionally responsive AI like Replika reveal diverse user reactions; some form stronger emotional attachments and deeper bond \cite{verma2023they, chen_wash2021, mou2017media}, while others report discomfort and feer with overly human-like behaviors \cite{li2024finding}.
This variability in emotional responses highlights a broader challenge within Human-Computer Interaction (HCI) research: understanding how systems can effectively evaluate or evoke empathy without causing discomfort or emotional dissonance. 
Empathy in AI presents a delicate balance, fostering emotional engagement while preserving authenticity. As emotional stakes rise, the difficulty of assessing and quantifying empathy in these systems grows, necessitating a more nuanced approach to evaluation \cite{cuadra2024illusion, chen2024feels}.
While human-centered design recognizes empathy as vital, measuring it within AI systems remains a major challenge due to its subjective, context-dependent nature \cite{chao2024multidisciplinary, zhu2024toward, dam2020empathy}.
To address this gap, this paper investigates how empathy is evaluated in AI versus humans and explores the factors that evoke empathetic responses in both. Through experimental methods using personal narratives, we examine how persona attributes, shared experiences, and model fine-tuning influence AI's alignment with human empathy judgments.
We also analyze how narrative qualities, such as emotional vividness and shared experiences and moral values, influence empathy evoked in AI and human responses.

\subsection{Empathy in the Psychological Literature}
The definition of empathy is multifaceted and encompasses a range of interpretations that highlight its significance in various fields such as healthcare, psychoanalysis, and interpersonal relations. Empathy is fundamentally described as the ability to understand and share the feelings of another, which is considered a crucial interpersonal skill \cite{davis2004empathy}. 
It involves both affective and cognitive components: emotional resonance with another's experience, and the deliberate understanding and communication of that experience, commonly framed as ``putting oneself in another's shoes'' \cite{peterson2017my, elliott2018therapist}.
In a relational framework, empathy is seen as an interactive and dynamic process where both the empathizer and the empathee shape each other's experiences, ultimately enhancing the quality of relationships \cite{van2020towards}. Experimental psychology identifies key factors influencing this exchange, such as the intensity and vividness of expressed emotion \cite{morelli2017empathy}, the perceived similarity between the empathizer and empathee \cite{krebs1975empathy, eklund2009ve}, and individual traits like gender, personality, and prior emotional experiences \cite{krebs1975empathy}.

When individuals engage in empathic social interactions, where they feel understood, supported, and valued by others, they experience a range of positive outcomes that contribute to overall well-being such as maintaining relationships \cite{reis2018intimacy}, and reducing stress, depression, and loneliness \cite{cohen1985stress, thoits2011mechanisms}.  
This aligns with psychoanalytic and therapeutic perspectives, which view empathy not just as a skill but as a method of deep interpersonal engagement and healing.
In mental health contexts, empathy fosters trust between patients and caregivers, enabling more effective support and treatment \cite{elliott2011empathy, elliott2018therapist}. 
According to \cite{miller2012motivational}, high-quality interactions occur when counselors focus on the client and show empathy, while low-quality interactions involve counselors giving instructions and the client merely complying. 
Empathetic communication is therefore essential in creating environments where individuals feel safe, heard, and emotionally validated—foundational for effective therapeutic and physician-patient relationships.

\subsection{Empathy in the Era of AI}
Within the HCI Community, empathy is recognized as a fundamental component of interpersonal and communication competence that augments understanding, prediction, persuasion, compliance gaining, relational development, and counseling among individuals \cite{redmond1989functions}. This view underscores the role of empathy not only in fostering effective human-to-human interactions but also in shaping more emotionally intelligent human-computer interactions. 
Recent advances in affective computing have enabled technologies to simulate emotional expressions and respond to users' affective states, paving the way for systems that can exhibit forms of empathy.

In the human-AI interaction domain, the role of AI in social and emotional contexts has been explored, making them suitable for applications such as social chatbots, where building rapport and sustaining engagement are essential. Some studies have indicated that empathetic AI agents are effective at fostering social connections, encouraging self-disclosure, facilitating social interactions, and maintaining user engagement \cite{bickmore2005establishing, devault2014simsensei, paiva2017empathy, zhou2020design}. However, their ability to fully replicate the nuanced empathy of humans remains limited. 
In therapeutic contexts, AI-driven counselors can offer consistent and accessible support but often fall short in emotional depth which is central to human empathy. 
Patients frequently express a preference for human-operated counselors, an agent controlled by AI, finding AI-driven empathy less helpful and sometimes counterproductive. However, providing attentive comments and offering hope have been shown to improve the perceived quality of AI-driven counseling, especially in emotionally charged contexts \cite{shao2023empathetic}. Furthermore, some studies have shown that AI agents can show biased value judgments and uneven expressions of empathy toward different demographic groups \cite{cuadra2024illusion, gabriel2024can}. These findings highlight the challenges AI faces in replicating human empathy, particularly given the complexity of empathy in emotionally sensitive and therapeutic settings, where shallow or biased responses can negatively affect therapeutic outcomes and .

To better understand these shortcomings, it is important to examine how empathy is evaluated and perceived differently in AI and human interactions.  Research on human-to-human interaction, such as online peer support communities that rely heavily on peer-based emotional support, indicated that highly empathetic responses are often rare \cite{cuadra2024illusion}. Through a mixed-methods analysis, \cite{syed2024machine} found that techniques like active listening and reflective restatements increased perceived empathy, while rigid structures and a lack of emotional validation diminished it. 
These findings suggest that effective empathy, either in human or AI, requires intentional strategies, not merely access to emotional content or communication channels.

Emerging technologies like large language models (LLMs) have opened new opportunities for simulating emotional processes more effectively. Deployed in social chatbots and mental health platforms, these systems show promise in boosting engagement and user satisfaction \cite{zhou2020design, alanezi2024assessing}.
Yet their long-term psychological impact and ethical implications remain underexplored, especially in intimate, support-focused roles. Empathy is also increasingly central in human-centered technology design, where understanding users' emotions and lived experiences guides product development \cite{rifat2024cohabitant,winters2021can,seeger2021texting}. This has led to empathetic design frameworks that use storytelling, cultural probes, and other immersive methods to help designers better see from users' perspectives \cite{debnath2024empathich}.

In virtual interactions with LLMs and conversational agents (CAs), the ability of virtual agents to appear more likable, trustworthy, and caring, underscores the substantial role of empathy in transforming the quality of human-AI interactions. Prior research has shown that empathetic characteristics can be modeled and embodied in virtual agents. For example, \cite{sharma2020computational} proposed a computational model of empathy, showing how the perception and understanding of others' emotional states can be algorithmically modeled and embedded within AI systems, and offering rich insights into the operationalization of empathy.

Despite these advances, a major challenge remains: how do we measure perceived empathy in technology? Unlike human empathy, which has well-established scales \cite{davis1983measuring, decker2014development}, the field lacks validated tools for evaluating empathy in AI systems. To address this, \cite{schmidmaier2024perceived} introduced the Perceived Empathy of Technology Scale (PETS)—a 10-item, 2-factor instrument that measures how empathetically users perceive interactive systems. This scale provides researchers and designers with a framework for evaluating the empathy exhibited by systems such as CAs and social robots \cite{schmidmaier2024perceived}. While CAs can simulate empathy \cite{paiva2017empathy,mcquiggan2007modeling}, they often fall short in truly interpreting users' emotional experiences, reinforcing the need for more nuanced, user-informed approaches to empathetic system design \cite{cuadra2024illusion}.

In light of these findings, the current study aims to deepen our understanding of how AI-generated empathetic responses differ from human judgments of empathy, and what factors influence the perception and evocation of empathy in both AI and humans. Our study is based on the previous research examining empathy through the heterogeneous effects of personal stories, which utilized a set of narrative stories in which individuals reflected on the three best and three worst events of their lives. These narratives were presented to Amazon MTurkers, who evaluated the extent of empathy they experienced and identified the key elements that evoked their empathic judgments. Accordingly, the following research questions formulated aim to address these critical gaps in the literature:

\begin{itemize}

\item RQ1. To what extent is empathy evaluated differently by humans and AI?

\item RQ2. How do various persona attributes—such as (a) gender, (b) empathic personality traits (empathic concern, perspective-taking), and (c) shared experiences with storytellers- influence the empathy assessed by AI, compared to humans? 

\item RQ3. To what extent does fine-tuning AI models improve the alignment between AI-generated and human-evaluated empathic responses? 

\item RQ4. What key factors influence the evocation of empathy in AI-generated responses compared to human responses? 
\end{itemize}

Using statistical analysis, we first compare how empathy is evaluated by AI, compared to humans. To make the AI-generated responses more human-like, we incorporate persona attributes into the prompts,  such as gender, empathic personality traits, and similarity of experience with the storyteller, drawing from psychological literature \cite{krebs1975empathy}. To further improve model performance, we applied instruction fine-tuning to GPT-4o. This was done in two ways: (1) fine-tuning on human-annotated empathy ratings associated with the story narratives, and (2) fine-tuning with additional reader attributes including gender, empathic concern, perspective-taking, and perceived similarity to the storyteller. We then quantitatively analyze the factors that evoke empathy in humans versus AI responses, emphasizing the focus on story attributes and the role of shared experiences in shaping empathic reaction  \cite{krebs1975empathy,eklund2009ve}. Finally, we highlight the need for thoughtful consideration of the potential benefits and harms of empathetic AI, particularly different effects across diverse subgroups and its implications in sensitive domains like mental health. By critically examining both the promise and the limitations of AI-mediated empathy, this study can contribute to the development of the next generation of AI systems that are not only technically proficient but also emotionally attuned and ethically aligned with human values and needs.

\section{Method}
\label{sec:method}

\subsection{Data}

\subsubsection{Human-Generated Data}
\begin{figure}
    \centering
    \includegraphics[width=0.5\linewidth]{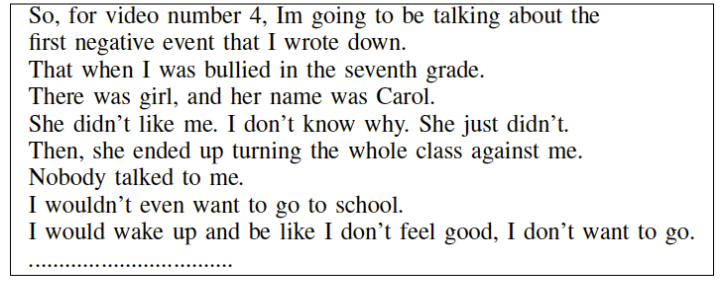}
    \caption{An example of the human-generated story used in the surveys}
    \label{fig:stories}
\end{figure}

Our analysis is based on data collected through an online survey conducted via Amazon MTurk in the winter of 2019 \cite{roshanaei2024paths}. This survey builds on previous IRB-approved research, where 756 videos of 126 undergraduate students were recorded. Participants were recruited using the Psychology Subject Pool and described the three best and three worst events of their lives. To ensure participant comfort with the use of their recorded videos in future studies, they have been asked to give consent for the videos in two steps. Upon arriving at the lab, participants completed a Pre-Video Recording consent form to provide consent to participate in the study. Then, after recording their videos, they filled out a Post-Video Recording consent form, explicitly indicating consent for the use of each recorded video separately. Later, each two-minute video was transcribed into text for analysis, with positive and negative labels assigned whether the story described one of the best or worst events of their life. In addition to these recorded videos, their demographic information (i.e., age,
gender, race) and personality characteristics of each participant, referred to hereafter as storytellers (see Figure \ref{fig:stories} for an example), were recorded. 

After compiling the narrative stories, a second IRB-approved study was conducted where participants were recruited via Amazon Mechanical Turk (MTurk). Participants were based in the U.S., aged 18 and older who had completed at least a high school education. All participants received monetary compensation for their time. The final sample included 2,586 individuals, with each narrative annotated by an average of three raters. Of the MTurk participants, 56\% identified as female, with an overall mean age of 38.6 years (SD = 12.58); the average age among male participants was 36.56 years (SD = 12.24).

During the survey, MTurk participants were asked to read a series of stories and respond to several questions using a 5-point Likert scale ranging from 1 (Not at all) to 5 (Extremely). These included assessments of: (1) overall empathy (e.g., To what extent did you feel empathy for the storyteller?), (2) the affective dimension of empathy, calculated as the average of responses to items assessing feelings of sympathy, compassion, and being moved, (3) the cognitive dimension of empathy, and (4) participants' reasons for the empathy they experienced toward each storyteller. A full list of survey items is provided in Supplementary Materials \ref{appendix:1}.
Given prior research highlighting gender differences in empathy \cite{sommerlad2021empathy, cohn1991sex, feingold1994gender}, as well as the established links between empathy evocation and personality traits related to empathy \cite{davis1980interpersonal, davis1983effects}, collecting demographic information (e.g., age, gender) along with measures of empathic concern and perspective taking, measured by Interpersonal Reactivity Index (IRI) \cite{davis1980interpersonal}.

In addition to individual traits, several studies have examined how the perceived relationship between the storyteller and the reader influences empathic responses, particularly the role of perceived similarity. For example, \cite{krebs1975empathy} found that readers who perceived greater similarity to a storyteller in terms of personality or values exhibited stronger physiological responses (e.g., increased heart rate and sweating) when exposed to the storyteller's pain.
To assess perceived shared experience in our study, two survey items were included: one measuring emotional similarity and another measuring similarity in the specific details of the experience. These items are described in full in Supplementary Materials \ref{appendix:1}.

\subsubsection{LLM-Generated Data}
To compare the levels of empathy perceived by AI versus humans, we used OpenAI's GPT-4o (gpt-4o-2024-08-06) \cite{achiam2023gpt} to assess the same human-generated stories that were previously analyzed by MTurk workers. To elicit comparable responses from the AI model, we developed a set of prompts based on the same questions posed to the MTurk workers (Supplementary Materials \ref{appendix:1}), structured in the following format:   
\begin{center}
Prompt = Instruction + Question + Output Format 
\end{center}

The example of a base prompt is added in Supplementary Materials \ref{appendix:2}. 

\subsubsection{LLM-Generated Data with Persona}
To answer RQ2, and evaluate the extent to which gender, IRI, and similarity of experience may influence the level of empathy AI experiences, we test four different treatment persona settings in total:

\begin{itemize}
\item AI-agent has gender. 
\item AI-agent has an empathic concern. 
\item AI-agent has perspective taking.  
\item AI-agent has an experience similarity with the storyteller. 
\end{itemize}

To implement these personas in the prompt, we used the same gender, empathic concern, perspective taking, and the level of similarity of experience of MTurk workers and embedded them in the prompts with the following structure: 
\begin{center}
Prompt = Instruction + Persona + Question + Output Format 
\end{center}

The example of a persona-based prompt is added in Supplementary Materials \ref{appendix:2}. 

\subsubsection{Fine-tuning Experiment}
We further explored methods of improving model performance using instruction fine-tuning to enable GPT-4o to perform more effectively. For the fine-tuning experiment, the prompt design was kept consistent with the methodology outlined above. 
Our data were stratified to create a balanced distribution across empathy scales, persona attributes (gender, empathic concern, perspective taking, and experience similarity), and positive and negative story tags, forming a candidate training set. Each story was treated as a single unit, with a composite categorical class constructed by concatenating the values of the stratified features. This approach ensures that stratification respects the joint distribution of these variables across training and test sets. Next, we counted the frequency of each composite class. Those with fewer than two occurrences were considered ``rare'' and handled separately, as they cannot be properly stratified. For non-rare classes, a stratified train-test split was applied using the composite class as the stratification criterion, and rare stories were split using a random (non-stratified) train-test split to ensure representation. Stories not assigned in either the stratified or rare splits were placed in the test set. Additionally, we checked stories in training and test sets, verified that no story appeared in both sets, and removed duplicates if necessary. Finally, we created a balanced instruction training set, comprising 80 stories with a similar distribution to our test set. Histograms of key categorical variables (e.g., tag, empathy, gender) were plotted to confirm that stratification preserved their distributions (see Supplementary materials Figure \ref{fig:stratify}). Additionally, Chi-square tests and L1-distance were used to quantify the similarity of distributions between the original and stratified datasets (see Supplementary materials Table \ref{tab:supp-s1}).
Fine-tuning is conducted in two nested steps: first, by using the stories, that is, the stories paired with corresponding human-annotated empathy ratings (GPT-4o FT Story-only). Second, building upon this, the model is further fine-tuned using not only the stories and empathy ratings but also additional reader-level attributes, including gender, similarity of experience with the storytellers, and self-reported empathic concern and perspective taking (GPT-4o FT All).

\subsection{Analytical Strategy}

To answer the first research question, We employed mean and standard deviation, correlation analysis, and t-test to evaluate the extent to which GPT-4o's empathy rating scores differ from human annotations. We also use
the Wasserstein distance, also known as the Earth Mover's Distance (EMD), to compare the probability distributions
between GPT-4o and human perceived empathy. 
Additionally, we applied the same evaluation metrics to assess the extent of differences in affective and cognitive dimensions of Empathy. We further use the same evaluation metrics that have been proposed for RQ1, to answer RQ2 (asses the impact of ``Persona'') and RQ3 (asses the impact of fine-tuning).   

To answer RQ4, and understand the extent to which factors evoke empathy in humans vs. AI, we used the Empathy reason questions including the situation of storytellers, and the extent of similarity with storytellers, both annotated by humans and GPT-4o. Research indicates the extent to which the storytellers' situations (e.g., intense) and sharing similar experiences play a role in evoking empathy \cite{morelli2017empathy, krebs1975empathy}. 
Due to the nested structure of our data in which each story is annotated with 2 to 4 humans, we used frequentist multilevel models, using the lme4 package \cite{bates2014fitting} in R version 4.4.1. We fit models with annotated data (Level 1) nested within stories (Level 2). We included a random intercept for each story, ICCs indicates the degree of variability between stories in our dependent variables (see Table \ref{tab:story_characteristics} and \ref{tab:story_similarity}). 
We centered and standardized our independent variables following recommendations for multilevel models \cite{curran2011disaggregation, yaremych2023centering}.  


\section{Results}
\label{sec:Results}

\subsection{RQ1: Empathy Alignment in Human and GPT-4o Vanilla}

\begin{figure}[htbp]
    \centering


        \includegraphics[width=\linewidth, height=0.25\textheight, keepaspectratio]{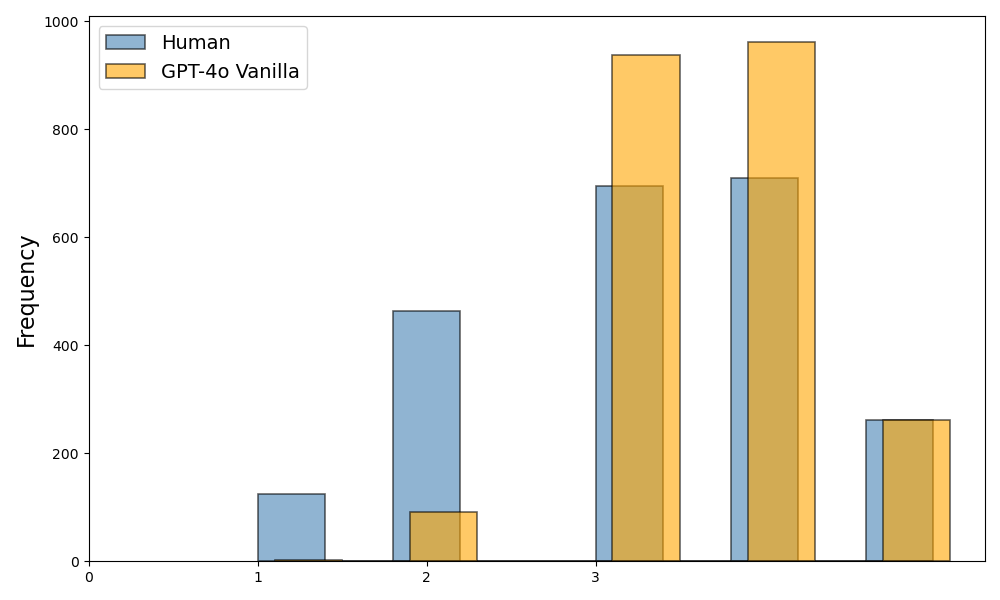} 
        \caption*{(a) Empathy}
        \label{fig:2-a}

    \vspace{0.5cm} 


        \includegraphics[width=\linewidth, height=0.25\textheight, keepaspectratio]{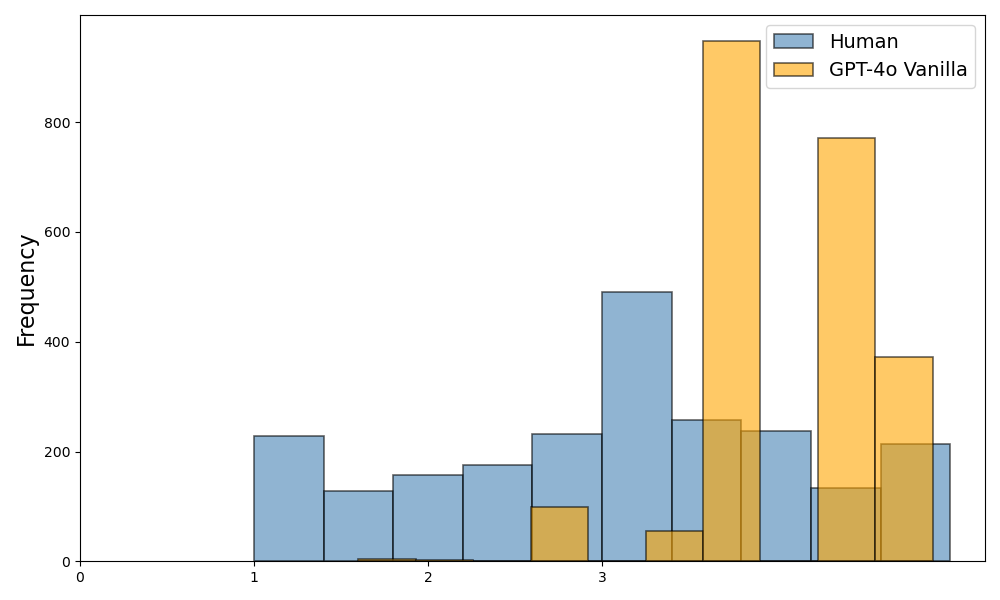} 
        \caption*{(b) Empathy-Affect}
        \label{fig:2-b}

    \vspace{0.5cm}


        \includegraphics[width=\linewidth, height=0.25\textheight, keepaspectratio]{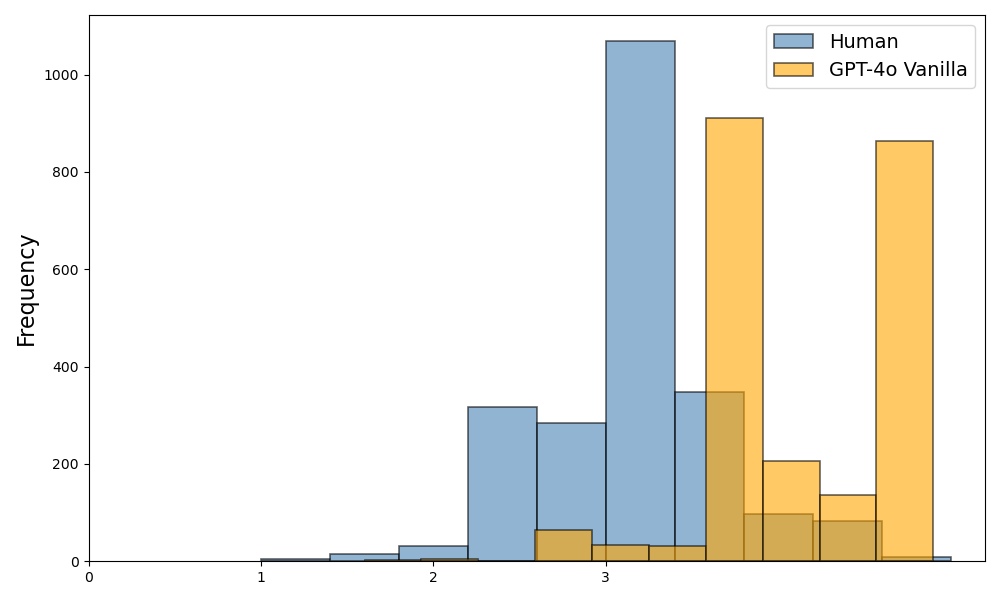} 
        \caption*{(c) Empathy-Cognition}
        \label{fig:2-c}

    \caption{Empathy Alignment in Human, compared to GPT-4o Vanilla}
    \label{fig:RQ1}
\end{figure}

We assessed how well LLM-generated ratings aligned with human judgments across three dimensions: overall empathy, emotional reactivity (affective empathy), and perspective-taking (cognitive empathy).
Our findings indicate that GPT-4o rates overall empathy higher with less variability compared to humans, for human: mean: 3.23, std: 1.074 and for GPT-4o: mean:3.615, std: 0.745, see Figure \ref{fig:RQ1} (a). This trend has been also observed in both dimensions of Empathy, affective dimension (human: mean: 3.049, std: 1.084, GPT-4o: mean: 3.893, std: 0.56), and cognitive dimension (human: mean: 3.085, std: 0.514, GPT-4o: mean: 4.001, std: 0.499), see Figure \ref{fig:RQ1} (b and c). 

\begin{table}[htbp]
    \centering
\resizebox{\textwidth}{!}{%
    \begin{tabular}{|c|c|c|c|c|}
        \hline
        & \textbf{Pearson(r,t,p-value)}  & \textbf{Cohen's~d} & \textbf{Wasserstein distance} & \textbf{t-test, p-value} \\ \hline
        \textbf{Empathy} & (0.25, 12.24,  0.001)   & -0.34 & 0.386 & (-15.996, p$<$.001)\\ \hline  
        \textbf{Empathy-Affect} & (0.35,16.316, 0.001) &  -0.81 &  0.866 & (-38.285, p$<$.001)  \\ \hline
        \textbf{Empathy-Cognition} & (0.019, 0.893, 0.372)  & -1.29 & 0.917 & (-61.277, p$<$.001) \\ \hline
    \end{tabular}}
    \caption{Empathy Alignment in Human, compared to GPT-4o Vanilla}
    \label{tab:RQ1}
\end{table}

As shown in Table \ref{tab:RQ1}, GPT-4o showed a significant moderate correlation, with human ratings of empathy, (r = 0.25, p$<$.001), with a small-to-medium effect size (Cohen's~d = -0.34, t = -15.996, p$<$.001), and significant differences in distribution (Wasserstein distance = 0.386). 

In the case of the affective dimension, we observed stronger correlation between human and AI (r = 0.35, p$<$.001), but the discrepancy between GPT-4o and human ratings increased (Cohen's~d = -.81, t = -38.285, p$<$.001,  Wasserstein distance = 0.866). These findings indicate a substantial overestimation of emotional reaction by the AI. Compared to affect, the weakest alignment was observed for the cognitive dimension of empathy, with a negligible correlation (r = 0.019, p$<$.372), with a big discrepancy between GPT-4o and human ratings (Cohen's~d = -1.29, t = -61.277, p$<$.001, Wasserstein distance = 0.917). These findings indicate a divergence in how GPT-4o understands cognitive empathy compared to human raters. Lacking lived experience, the model cannot fully relate to human stories or communicate an understanding of users' feelings in a meaningful way. Therefore, while GPT-4o can partially approximate human judgments of affective empathy,  it struggles with the deeper and more nuanced aspects of cognitive empathy. 

\subsection{RQ2: Empathy Alignment in Human and GPT-4o Vanilla with Persona}
We further examined how different treatment personas, a) gender, b) empathic concern, c) perspective taking, and d) similarity of experience, impact the empathy ratings. As shown in Table \ref{tab:RQ2-persona}, adding gender slightly decreased correlation with human ratings (r = 0.232, p$<$.001), and increased the difference between GPT-4o and human ratings (Cohen's~d = -0.401, t = -19.047, p$<$.001, Wasserstein distance = 0.462). This finding suggests that gender persona alone might not be effective in improving alignment. In fact, the overall performance slightly worsens compared to the vanilla (base) model.

Looking into the results of other persona attributes, our findings indicate slightly more correlation (less variability) with human empathy (EC: r = 0.318, p$<$.001, PT: r = 0.286, p$<$.001, Sim: r = 0.346 p$<$.001); However, their discrepancy increased (EC: Cohen's~d = -0.378, t = -17.96, p$<$.001, Wasserstein distance = 0.436, PT: Cohen's~d = -0.385, t = -18.302, p$<$.001, Wasserstein distance = 0.484, Sim: Cohen's~d = -0.458, t = -21.78, p$<$.001, Wasserstein distance = .541). The results indicate more divergence in the mean difference and distribution, suggesting that incorporating persona attributes leads to ratings that are directionally aligned with human judgments, however notable divergences from human empathy ratings still persist.
 
\begin{table}[htbp]
    \centering
    \resizebox{\textwidth}{!}{%
    \begin{tabular}{|c|c|c|c|c|c|c|}
        \hline
        &\textbf{Model}& \textbf{Pearson(r,t,p-value)} & \textbf{Cohen's~d} & \textbf{Wasserstein distance} & \textbf{t-test, adjusted p-value}  \\ \hline
        
         \multicolumn{1}{|c|}{}& \textbf{GPT-4o Vanilla} 
         & (0.250,12.235, p$<$.001)	 &	-0.337 & 0.386	& (-15.99,p$<$.001)	 \\  
        \multicolumn{1}{|c|}{} &\textbf{GPT-4o Vanilla with Gender}  
        &(0.232, 11.315, p$<$.001)	 &	-0.401 & 0.462 & (-19.05,p$<$.001)  \\ 
        Empathy&\textbf{GPT-4o Vanilla with Empathic Concern} 
        &(0.318, 15.893, p$<$.001 )	 &	-0.378 & 0.436 & (-17.96,p$<$.001)   \\ 
        \multicolumn{1}{|c|}{} &\textbf{GPT-4o Vanilla with Perspective Taking}  
       &(0.286, 14.144, p$<$.001)	 &	-0.385 & 0.484 & (-18.302,p$<$.001)  	\\ 
         \multicolumn{1}{|c|}{}&\textbf{GPT-4o Vanilla with Experience Similarity}  
         & (0.346,17.473, p$<$.001)	 &	-0.458  & 0.541 & (-21.78,p$<$.001)	  \\ \hline
    \end{tabular}}
        \caption{Empathy Alignment in Human, compared to GPT-4o Vanilla with Persona Included in Prompt}
    \label{tab:RQ2-persona}
\end{table}

\subsection{RQ3: Empathy Alignment in Human and GPT-4o Using Fine-Tuned Models}
We also examined the extent to which fine-tuning GPT-4o influences its alignment with human empathy ratings, using two steps. In the first step, fine-tuning was conducted using the training set of 80 stories along with human-annotated empathy ratings. In the following step, we incorporated reader-reported attributes including gender, empathic concern, perspective taking, and similarity of experience with storytellers, into the fine-tuning process. 

As shown in Table \ref{tab:RQ3}, fine-tuning only on the story content (FT Story-only), decreased the correlation slightly (r = 0.213), however, the discrepancy also decreased, leading to stronger alignment (Cohen's d = -0.288, t = -13.684, p$<$.001, Wasserstein distance = 0.494). This result indicates that focusing solely on narrative content improves the performance of AI. While the model fine-tuned on both stories and user attributes (FT All) showed a similar correlation (r = 0.219), it drastically decreased its divergence from humans (Cohen's d = 0.007, t = 0.33, p$<$.74, Wasserstein distance drops to 0.324), indicating a near-complete alignment in overall empathy distribution.

\begin{figure}[htbp]
    \centering


        \includegraphics[width=\linewidth, height=0.25\textheight, keepaspectratio]{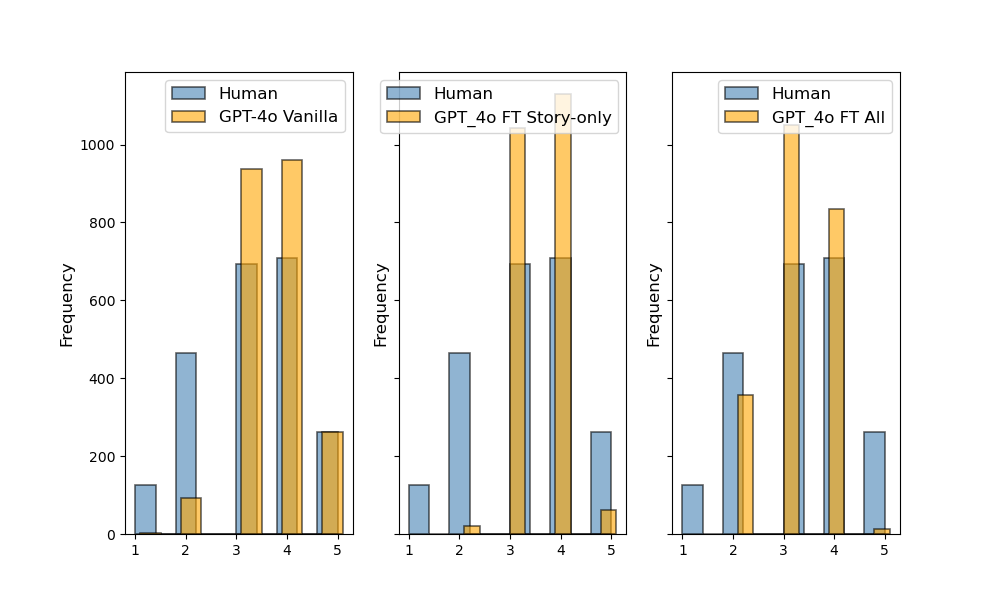} 
        \caption*{(a) Empathy}
    \vspace{0.5cm} 

        \includegraphics[width=\linewidth, height=0.25\textheight, keepaspectratio]{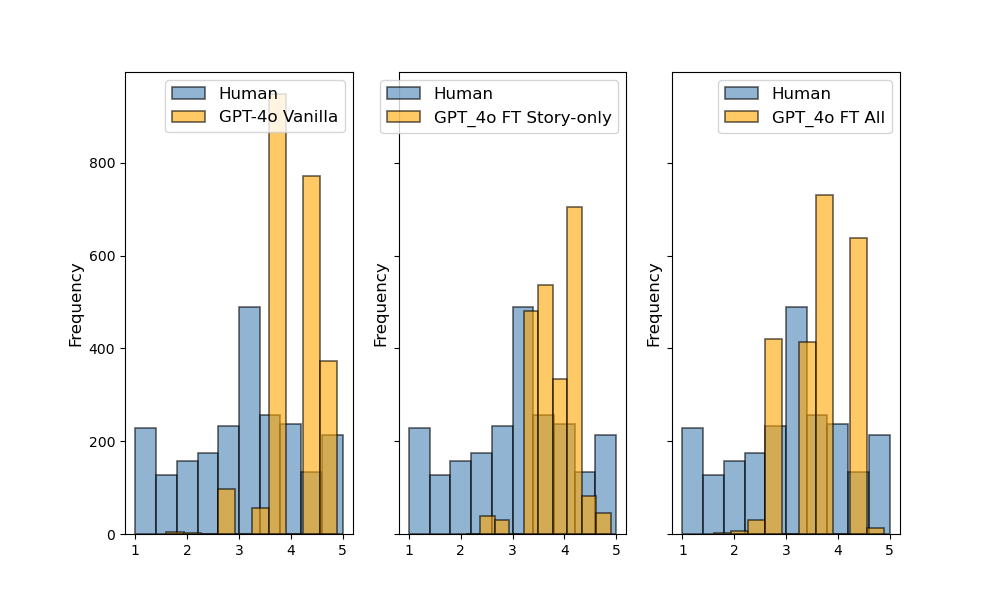} 
        \caption*{(b) Empathy-Affect}
    \vspace{0.5cm} 
    
        \includegraphics[width=\linewidth, height=0.25\textheight, keepaspectratio]{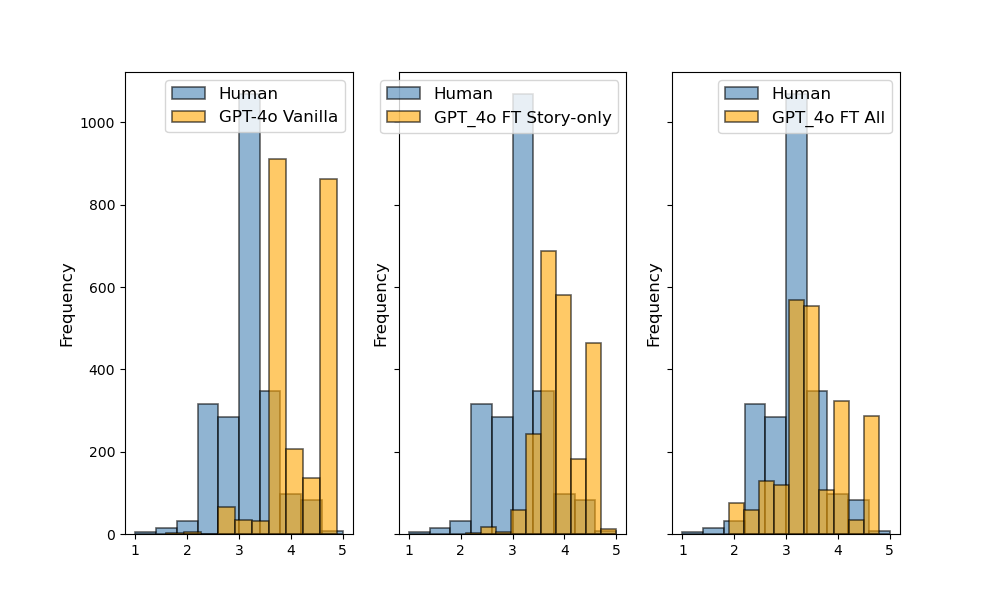} 
        \caption*{(c) Empathy-Cognition}
    \vspace{0.5cm}
    
    \caption{Empathy Alignment in Human, compared to GPT-4o Vanilla and the Fine-Tuned Models (Story Only and All Attributes)}
    \label{fig:RQ3}
\end{figure}

Similar to overall Empathy, we observed improvement in alignment in both affective and cognitive dimensions of empathy (see Table \ref{tab:RQ3}). In particular,  the discrepancy dropped significantly in cognition, while distributional alignment improved (Cohen’s d = -0.467, t = -22.18, p$<$.001, Wasserstein distance = 0.40). This result suggests that adding user context while fine-tuning helps GPT-4o better approximate human cognitive empathy (Figure \ref{fig:RQ3}(c)). 

\begin{figure}
    \centering
    \includegraphics[width=0.5\linewidth]{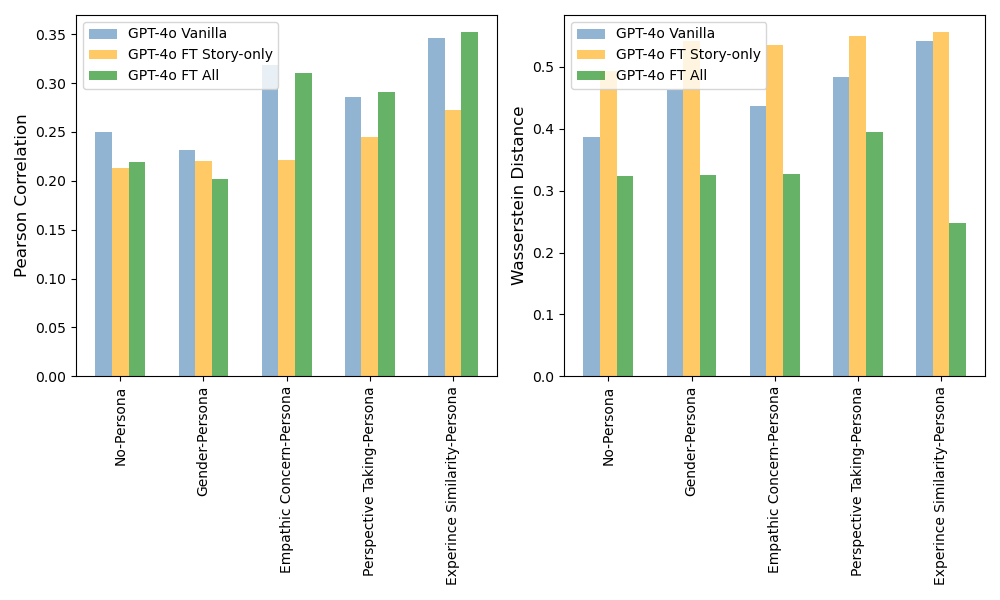}
    \caption{Empathy Alignment in Human, compared to GPT-4o Vanilla and Fine-Tuned Models Pearson Correlation (left) and Wasserstein Distance (right)}
    \label{fig:RQ3-comparision}
\end{figure}

Our previous results showed that incorporating personas into GPT-4o prompts can only slightly boost performance in AI, compared to human empathy ratings. However, including this information to prompts after fine-tuning GPT-4o on both stories and human attributes, significantly improved models' empathy alignment and made its responses more human-like, see Table \ref{tab:RQ3-persona}, but no significant difference in correlation results was observed. After fine-tuning, the strongest impact was achieved when the persona attribute reflected the percived shared experience with the storytellers, see Figure \ref{fig:RQ3-comparision}. As shown in Table \ref{tab:RQ3-persona}, the highest correlation (0.35, p$<$.001), lowest Wasserstein distance (0.248), with minimal mean difference (Cohen's d = -0.049, t = -2.323, p$<$.02), indicating the closest match in distribution and smallest mean divergence was achieved between Human ratings and fine-tuned model (GPT-4o FT All).  
Overall, incorporating persona attributes after fine-tuning led to aligning the model responses with human judgments, indicating that tailoring the model to better reflect the user's persona can enhance its performance in generating empathetic or relevant responses

\begin{table}[htbp]
    \centering
    \resizebox{\columnwidth}{!}{%
    \begin{tabular}{|c|c|c|c|c|c|}
        \hline
        \textbf{}&\textbf{Model}& \textbf{Pearson(r,t,p-value)} & \textbf{Cohen's~d} & \textbf{Wasserstein distance} & \textbf{t-test, adjusted p-value}  \\ \hline
         \multicolumn{1}{|c|}{ }& \textbf{GPT-4o Vanilla}& (0.25, 12.235, p$<$.001)	 &  -0.337 &	0.386 & (-15.99, p$<$.001)	 \\ 
        \multicolumn{1}{|c|}{\multirow{3}{*}\textbf{Empathy}} &\textbf{GPT-4o FT Story-only} & 
        (0.213, 10.366, p$<$.001)		&-0.288&	0.494 &
        (-13.684, p$<$.001) \\  
         \multicolumn{1}{|c|}{ }&\textbf{GPT-4o FT All} &(0.219, 10.650, p$<$.001 )		& 0.007&	0.324 & (0.33, p$<$.74) \\ \hline
        \multicolumn{1}{|c|}{ } & \textbf{GPT-4o Vanilla} 
        & (0.325, 16.316, p$<$.001)	&-0.806	& 0.866 & (-38.29, p$<$.001)	 \\  
        \multicolumn{1}{|c|}{\multirow{3}{*}\textbf{Empathy-Affect}} &\textbf{GPT-4o FT Story-only} 
        & (0.317, 15.842 , p$<$.001)	&-0.66	&0.784 & (-31.30, p$<$.001) \\
        \multicolumn{1}{|c|}{ } &\textbf{GPT-4o FT All} 
        & (0.313, 15.635 , p$<$.001 )	&-0.403	&0.565 & (-19.17 , p$<$.001) \\ \hline
        \multicolumn{1}{|c|}{ } & \textbf{GPT-4o Vanilla} 
        &(0.019, 0.893, p$<$.372)	&-1.29	&0.917 & (-61.28, p$<$.001 )\\ 
        \multicolumn{1}{|c|}{\multirow{ 3}{*}\textbf{Empathy-Cognition}} &\textbf{GPT-4o FT Story-only} 
        &(0.010, 0.940, p$<$.347)		&-1.052	&0.73 & (-49.98,p$<$.001 )	 \\ 
        \multicolumn{1}{|c|}{ } &\textbf{GPT-4o FT All} 
        &(0.020,0.488, p$<$.626 )		&-0.467	&0.40 & (-22.18,  p$<$.001 )
        \\ \hline		 
    \end{tabular}}
        \caption{Empathy Alignment in Human, compared to GPT-4o Vanilla and the Fine-Tuned Models (Story Only and All Attributes)}
    \label{tab:RQ3}
\end{table}

\begin{table}[htbp]
    \centering
    \resizebox{\textwidth}{!}{%
    \begin{tabular}{|c|c|c|c|c|c|c|}
        \hline
        &\textbf{Model}& \textbf{Pearson(r,t,p-value)} & \textbf{Cohen's~d} & \textbf{Wasserstein distance} & \textbf{t-test, adjusted p-value}  \\ \hline
        
         \multicolumn{1}{|c|}{}& \textbf{GPT-4o FT All} 
         & (0.219, 10.650, p$<$.001)	  & 	0.007	 &   0.324 & (0.330, p$<$.74)	 \\  
        \multicolumn{1}{|c|}{} &\textbf{GPT-4o FT All with Gender}  
        & (0.202, 9.800, p$<$.001)	  & 	-0.054	 &  0.326	 &  (-2.587, p$<$.01)  \\ 
        \multicolumn{1}{|c|}{\multirow{5}{*}\textbf{Empathy}}&\textbf{GPT-4o FT All with Empathic Concern} 
        & (0.310, 15.486, p$<$.001)	  & 	-0.113	 &  0.327	 &  (-5.385,p$<$.001)  \\ 
        \multicolumn{1}{|c|}{} &\textbf{GPT-4o FT All with Perspective Taking}  
        & (0.291, 14.444, p$<$.001)	  & 	-0.165	 &  0.394	 &  (-7.864, p$<$.001) 	\\ 
         \multicolumn{1}{|c|}{}&\textbf{GPT-4o FT All with Experience Similarity}  
         & (0.352, 17.856, p$<$.001)	  &  	-0.049	 &  0.248	 &  (-2.323, p$<$.02)	  \\ \hline
    \end{tabular}}
        \caption{Empathy Alignment in Human, compared to GPT-4o Fine-Tuned Model (All Attributes) with Persona Included in Prompt }
    \label{tab:RQ3-persona}
\end{table}

\subsection{RQ4: Empathy Evocation in Human and GPT-4o} 
To better understand the differences in empathy ratings between humans and GPT-4o, we examined how the story's characters and perceived similarity across various dimensions may influence the degree to which humans and GPT-4o empathize with the story. From Table \ref{tab:story_characteristics} qualities of the story itself predict empathy, comparing human and GPT conditions. Emotionally rich and vivid stories evoked significantly more empathy overall (Emotional: B: 0.206,  p$<$.001, Vivid: B: 0.094, p$<$.001). Notably, GPT-4o outperformed human responses in these dimensions (Emotional: B: 0.103, p$<$.01, Vivid: B: 0.070, p$<$.05,  Dramatic: 0.123, p$<$.001), indicating GPT-4o is more sensitive to Emotional and dramatic stories. In contrast, stories described as exciting (B: -0.117, p$<$.001), or those with characteristics resembling real-life experiences (B: -0.068, p$<$.05 ), were less effective or even slightly counterproductive in evoking empathy in the GPT condition.

Our results in Table \ref{tab:story_similarity} show that perceived similarity between the reader and the storyteller significantly influences the reader's empathy. Specifically, empathy was higher when readers felt emotionally similar to the storyteller (B = 0.119, p $<$ .001), perceived similarity in story details (B = 0.079, p $<$ .001), or felt aligned in personality or moral values (Personality: B = 0.063, p $<$ .01). While GPT-4o evoked higher overall empathy, it showed reduced sensitivity to those similarity cues compared to human listeners(Emotional = -0.127, p$<$.001; Detail = -0.133, p$<$.001, Personality:-0.117, p$<$.001, Moral:-0.064, p$<$.05 ). Moreover, perceived demographic similarity had no significant effect on empathy in either the human or GPT-4o condition. 

\begin{table}[htbp]
\centering
\begin{tabular}{@{}lr@{}}
\toprule
\textbf{Variable} & \textbf{B\_Estimate} \\
\midrule
(Intercept) & 3.231 *** \\
GPT & 0.386 *** \\
Story\_Vivid\_GM & 0.094 *** \\
Story\_Abstract\_GM & 0.017 \\
Story\_Exciting\_GM & 0.045 * \\
Story\_EasytoImagine\_GM & 0.053 ** \\
Story\_Social\_GM & -0.034 \\
Story\_Personal\_GM & 0.069 *** \\
Story\_Coherent\_GM & -0.028 \\
Story\_Unpredictable\_GM & 0.018 \\
Story\_Emotional\_GM & 0.206 *** \\
Story\_Logical\_GM & 0.009 \\
Story\_Relevant.to.my.life\_GM & 0.081 *** \\
Story\_Dramatic\_GM & 0.038 \\
GPT×Story\_Vivid\_GM & 0.070 * \\
GPT×Story\_Exciting\_GM & -0.117 *** \\
GPT×Story\_EasytoImagine\_GM & -0.036 \\
GPT×Story\_Social\_GM & 0.007 \\
GPT×Story\_Personal\_GM & -0.034 \\
GPT×Story\_Coherent\_GM & 0.037 \\
GPT×Story\_Unpredictable\_GM & -0.051 \\
GPT×Story\_Emotional\_GM & 0.103 ** \\
GPT×Story\_Logical\_GM & -0.021 \\
GPT×Story\_Relevant.to.my.life\_GM & -0.068 * \\
GPT×Story\_Dramatic\_GM & 0.123 *** \\
\midrule
\multicolumn{2}{l}{\textbf{Random Effects}} \\
$\sigma^2$ & 0.61 \\
$\tau_{00}$ (text) & 0.21 \\
ICC & 0.26 \\
N (text) & 668 \\
\bottomrule
\end{tabular}
\caption{Linear Mixed Model Results on Empathy Evocation (a) Story Characteristics}
\label{tab:story_characteristics}
\end{table}

\begin{table}[htbp]
\centering
\begin{tabular}{@{}lr@{}}
\toprule
\textbf{Variable} & \textbf{B\_Estimate} \\
\midrule
(Intercept) & 3.231 *** \\
GPT & 0.386 *** \\
Sim\_Emotional & 0.119 *** \\
Sim\_Detail & 0.079 *** \\
Sim\_Age & -0.008 \\
Sim\_Gender & -0.021 \\
Sim\_Race & -0.020 \\
Sim\_Personality & 0.063 ** \\
Sim\_Moral & 0.111 *** \\
GPT×Sim\_Emotional & -0.127 *** \\
GPT×Sim\_Detail & -0.133 *** \\
GPT×Sim\_Age & -0.001 \\
GPT×Sim\_Personality & -0.117 *** \\
GPT×Sim\_Moral & -0.064 * \\
\midrule
\multicolumn{2}{l}{\textbf{Random Effects}} \\
$\sigma^2$ & 0.59 \\
$\tau_{00}$ (text) & 0.09 \\
ICC & 0.13 \\
N (text) & 668 \\
\bottomrule
\end{tabular}
\caption{Linear Mixed Model Results on Empathy Evocation (b) Storyteller Similarity}
\label{tab:story_similarity}
\end{table}

\section{Discussion}
Empathy plays a central role in Human-centered AI, influencing both the design process and the creation of tools and technologies that shape user experiences and interactions. Advancements in AI technology, in particular LLMs, have created new opportunities for designing systems that can more effectively emulate human emotional processes and enhance their ability to engage in empathetic interactions. Despite these advancements, the effectiveness and psychological impacts of AI-driven empathetic interactions remain under-explored. To address this, it is crucial to investigate how empathy is evaluated in human versus AI interactions, as well as the factors that evoke empathy in each of these settings.

In this work, we compared GPT-4o's ability to evaluate empathy vs. humans and examined how incorporating persona in prompts can improve the alignment of AI's empathetic responses with human judgments. We also explored the impact of fine-tuning the model to enable GPT-4o to perform more effectively. Finally, we examined factors that evoke empathy, including story attributes, and the degree of similarity with the storyteller, in both humans and AI. This exploration is crucial for designing the next generation of AI systems that are not only technically proficient but also emotionally intelligent and ethically aligned with human values and needs.

Our findings indicate overrated empathy by LLMs with less variability compared to Humans. While the model may simulate emotional response, it struggles to capture and convey the deeper, and context-dependent aspects of human experience, supporting previous findings \cite{cuadra2024illusion}. Therefore, its empathetic responses can appear overly intense or unrealistic, often beyond what the situation or emotional context requires. This can give users the impression of being heard, yet the mismatch with human norms may make interactions feel inauthentic. Furthermore, machines may be able to recognize and respond to human emotions in ways that feel socially appropriate and empathetic, however, the lack of authentic human presence, the act of offering time, being fully attuned to another's emotional state, is something machines can only simulate, not genuinely experience \cite{picard2000affective, turkle2011alone}. This limitation of current LLMs becomes especially concerning during moments of loneliness or emotional vulnerability when individuals are more likely to use AI as companions or friends. According to Epley's three-factor theory of anthropomorphism, the desire for social connection increases the tendency to attribute human-like characteristics to non-human agents \cite{epley2007threefactor}. In such moments, lonely individuals may be more susceptible to overreliance on AI, which can increase the risks of miscommunication, distorted expectations, and emotional harm. 

From the studies in psychology, the reader's characteristics such as gender, personality, age, and past experiences influence vicarious emotions \cite{krebs1975empathy}. Our results indicate that adding persona attributes to GPT-4o slightly decreases variability (more correlation) with human empathy ratings, but also increases discrepancies in how those ratings are distributed. The greatest improvements in alignment came from empathic concern and experience similarity, but even these showed significant differences in mean ratings and distribution compared to human judgments.   

While fine-tuning GPT-4o enhances its alignment with human responses, human empathy is inherently a multifaceted process that goes beyond a singular emotional or cognitive dimension. It involves consistency and a deeper understanding of emotional context, and it is about ``putting oneself in another shoes'' to fully grasp another person's perspective and feelings \cite{peterson2017my}. Fine-tuning allows the model to better align its responses with these complex human dynamics, reducing the unrealistic or artificial aspects of AI interactions and making its behavior more natural and relatable. Focusing solely on stories, the fine-tuned response is more human-like, while incorporating all readers' traits in the fine-tuning process. Including the relevant attributes (e.g., personality, gender, or experiences), decreases the model's discrepancy, enhances the model's 
consistency with human judgments, and allows the model to generate more nuanced and personalized responses and exhibit more human-like qualities in both emotional reactions and understanding.
Later, we additionally tested the influence of persona in a fine-tuned model and observed the biggest gains come when the model is tuned to reflect experience similarity of experience. A deeper understanding of each other's emotions and perspectives plays a critical role in empathy. Our findings indicated that contextual and personalized fine-tuning is key to enhancing AI empathy, leading the model to be better able to mimic this natural human connection and generate more realistic empathetic responses.

Experimental studies in psychology have indicated several factors that play a critical role in evoking empathy. For instance, readers tend to empathize more deeply when they share similar life experiences with the storyteller \cite{krebs1975empathy,eklund2009ve}, in particular when the storyteller's situation is particularly emotionally intense and vivid \cite{morelli2017empathy}, or the story touches on relatable domains such as work or home life \cite{roshanaei2024paths}. Our findings show that AI-generated responses often align with existing literature by producing higher overall empathy ratings. However, there are key factors where AI responses diverge from those of humans. AI-generated responses indicate less sensitivity to personal similarity cues, such as shared emotions, experiences, or moral values, while humans show more empathy when they feel emotionally or morally aligned with the storyteller. This highlights a key limitation of current LLMs with their lack of understanding of others' emotions and perspectives, which is essential for authentic empathic engagement. 

Our findings indicate that GPT-4o's empathy is significantly influenced by the emotional richness, vividness, and drama of the story itself, while it fails to empathize when exciting events occur. This pattern mirrors human tendencies but still indicates a larger discrepancy between the model and human empathy. To further validate these findings, we examined GPT-4o's performance separately for positive and negative stories (see Supplementary Materials \ref{appendix:4}). The model demonstrated a stronger correlation with human empathy ratings in negative contexts; however, it also tended to amplify the emotional intensity in these situations. In positive contexts, there was less alignment with human judgments, and the degree of overestimation was smaller. 
According to \cite{morelli2014neural, morelli2015common} there is an essential distinction between negative and positive empathy, which involves sharing differently valenced emotions (i.e., positive versus negative) and activating distinct regions in neural systems (e.g., ventromedial prefrontal cortex vs. anterior insula,
dorsal anterior cingulate cortex). 
Negative empathy, which is often referred to as empathy in classical models, refers to resonating with another's suffering and experiencing empathic concern and distress. In contrast, positive empathy reflects the ability to share and respond to others' positive emotions, such as joy, pride, or success, with feelings of vicarious happiness or celebratory concern \cite{morelli2015emerging}. Negative empathy typically motivates helping and supporting behaviors, whereas positive empathy supports relationship building, emotional closeness, and prosocial behavior \cite{batson1991empathic, smith1989altruism}.

Our finding shows that humans are capable of empathizing in both positive and negative contexts, like joy and excitement, as well as distress, whereas GPT-4o's responses reveal a marked gap in its ability to empathize with positive circumstances. This asymmetry may suggest a potential bias, where LLMs may overemphasize vivid, emotionally charged distress while under-responding to positive affect.
Prior research shows that people reveal greater positive empathy toward close others or in-group members, compared to negative empathy, which reflects concern and compassion, and tends to extend more broadly even toward out-group members \cite{morelli2015emerging}. Compared to humans, GPT-4o's expression of empathy appears more universal and impartial, yet it lacks the nuanced warmth and relational depth that characterize human empathy within close relationships. This pattern highlights GPT-4o's struggle to convey the intimacy, companionship, and shared joy that underpin positive empathy in human social relationships. This limitation raises important considerations for the design and deployment of empathetic AI systems, particularly in sensitive contexts such as counseling and emotional support.

Beyond emotional resonance, our findings also underscore that cognitive empathy, the capacity to infer another person's thoughts, intentions, and situational perspective, remains a core challenge for GPT-4o. Prior work in affective computing and AI ethics has noted that while such systems can imitate emotional expression, they struggle with perspective-taking and theory-of-mind reasoning, often simulating empathy rather than genuinely understanding it \cite{mitchell2019artificial,bender2021stochastic,marcus2020gpt3}. Unlike emotional empathy, which reflects affective alignment, cognitive empathy requires reasoning about others' beliefs and intentions, abilities closely tied to the theory of mind \cite{frith2005theory}. Since GPT-4o lacks embodied experience, long-term memory continuity, and grounded world models, its perspective inference remains based on linguistic associations rather than genuine comprehension. Future research could examine these alignment patterns more deeply to clarify what ``empathy'' entails across distinct systems of mind, human and artificial, and how each conceptualizes and interprets others’ emotions and intentions.

\subsection{Emotional and Social Implications}
The integration of empathetic LLMs into our lives holds significant emotional and social implications, and could potentially cause harm which we will discuss in this section. 

On an emotional level, empathetic AI would be able to enhance emotional well-being by offering accessible, and non-judgmental support for individuals who face barriers to human interaction due to geographical distance, and mental health issues, which are increasingly prevalent among college students \cite{beiter2015prevalence, mofatteh2021risk,pratt2000facilitating, kroencke2023well}. 
However, there are risks associated with relying on AI for emotional support. Building on our findings, even though AI can get better at mimicking empathy, it lacks the depth of authentic human understanding, potentially leading to superficial or even harmful responses in complex emotional situations \cite{li2023feasibility}. This raises concerns about users forming inauthentic emotional attachments with AI, which may lead to a sense of detachment or disillusionment, as noted by studies indicating that overly human-like AI behavior can provoke discomfort or fear \cite{li2024finding}.

Moreover, in the context of anxiety disorders, the use of empathetic AI systems may inadvertently serve as a form of safety behavior \cite{helbiglang2010safety}. safety behaviors are actions taken to avoid, escape, or minimize anxiety-provoking situations, offering temporary relief, however, they often maintain or even worsen anxiety over time. For instance, someone with social anxiety might avoid social events or conversations to escape feelings of discomfort or judgment. Although this avoidance reduces anxiety at the moment, it prevents individuals from confronting their fears and reinforces the belief that social situations are dangerous. In this situation, individuals with social anxiety might increasingly rely on AI for emotional support or social interaction, further avoiding real-life engagements and missing opportunities to develop coping strategies. This over-reliance on AI could ultimately hinder recovery and contribute to the persistence of anxiety symptoms \cite{helbiglang2010safety}.  

In broader social contexts, empathetic AI would be able to reshape interactions by affecting how people perceive sincerity and connection in their relationships. As these systems become more integrated into everyday communication, they may create new norms around digital communication and emotional intimacy \cite{mou2017media}. However, the use of empathetic AI in sensitive domains such as education and healthcare must emphasize augmentation rather than replacement of human interaction, ensuring that AI acts as a supplementary tool rather than a substitute for genuine human relationships \cite{lee2021social, li2023feasibility}. 

Despite these challenges, empathetic AI has the potential to positively improve mental health and social well-being, though it requires careful design to avoid reinforcing harmful biases and ensure ethical and effective emotional support systems.
\label{sec:discussion}

\section{Limitations and Future Work}
\label{sec:limitations and future work}
Our work investigates how empathetic responses generated by AI align with human empathy, using story-based scale ratings. One key concern would be a conflation between the measurement and the expression of empathy in this study. When AI is evaluated using rating scales, it may appear empathetic because it simply mimics surface-level cues, rather than genuinely understanding or sharing emotional perspectives. Using this system may oversimplify complex human emotions, and lead to formulaic AI responses that fail to foster genuine emotional engagement \cite{paiva2017empathy, elliott2011empathy}. Moreover, traditional rating metrics may fail to capture the subjective and context-dependent nature of empathy, especially when applied to AI systems. In addition, AI-generated responses are based on patterns learned from training data, which may result in different response distributions, and exhibit systematic biases, such as a tendency to provide higher ratings or more positive feedback, regardless of the content. In other words, it is possible that the AI's skewed responses are a characteristic of its general output style rather than a reflection of its empathetic capabilities. In another word, the inherent probabilistic nature of LLMs poses challenges in ensuring consistency, interpretability, and authenticity in their empathetic responses.
All these together, highlight the need to rethink how empathy is assessed and to develop more sophisticated models that can handle the complexity of human emotions. While our work complements works such as \cite{Shen2024}, more research is needed to understand the underlying factors driving these distinctions. To address these limitations, future research should explore qualitative assessments of empathy, focusing on the words, texts, and expressions used rather than relying solely on numerical scales. One key direction for future exploration is using the word attention mechanisms. By using attention-based architectures, we can investigate how LLMs interpret and prioritize specific words or phrases when generating empathetic responses, and how this differs from human empathy in similar settings. This could offer deeper insight into the cognitive and emotional processes underlying empathy in both humans and AI, and how different humans and AI perceive, process, and express empathy.

A second major concern in our work is the issue of personalization in empathetic AI systems. In this study, we examined the impact of incorporating persona (gender, personality, or shared experiences) and fine-tuning the models \cite{krebs1975empathy, eklund2009ve}. While personalization can improve the perceived empathy in AI systems, it also raises important ethical concerns, especially in healthcare. Healthcare data is highly sensitive, and the use of personal information to create realistic empathetic interactions requires strong safeguards to prevent misuse or breaches \cite{yalccin2019evaluating}. Moreover, AI systems must be carefully designed to avoid reinforcing existing biases in healthcare. For instance, if AI is trained using biased data, it could lead to disparities in how empathy is expressed toward patients from different demographics or cultural groups, potentially worsening existing inequalities in healthcare \cite{paiva2017empathy}. Given these concerns, future research must emphasize the urgent need for ethical guidelines and oversight in the development of empathetic AI, especially when personalized interactions are involved  \cite{yalccin2019evaluating}.

In addition, future research should expand the cultural and demographic diversity of participants to strengthen the external validity and generalizability of the findings. 
Because all participants in the current study were based in the United States, the observed patterns of empathy may reflect Western and individualistic norms surrounding emotional expression. Prior research has shown that various emotional experiences by individuals (e.g., anger, shame, empathy) are culturally shaped, varying across societies that emphasize collectivism versus individualism \cite{markus1991culture, boiger2018beyond, eichbaum2023empathy}. For example, some specific aspects of self-compassion and empathy, such as Self-Kindness, Common Humanity, and Isolation, vary significantly across cultures \cite{Birkett2013SelfcompassionAE}. Similarly, AI models trained mostly on Western-centric data may unintentionally encode cultural biases, which can limit their accuracy and fairness across diverse populations \cite{bender2021stochastic}. In future studies,  including participants from a broader range of cultural, linguistic, and socioeconomic backgrounds would help ensure that both human and AI empathy patterns are evaluated in a more globally representative context. Furthermore, the design of empathetic AI must account for inclusivity and adaptability across diverse cultural contexts, ensuring systems can interact with a wide range of emotional expressions and cultural norms. 

Finally, our study relies solely on GPT-4o and a single set of prompts, potentially constraining our understanding of how different language models respond to varying inputs. Future research should explore multiple prompt variations across a broader range of models to enable a more comprehensive analysis of model behavior. Additionally, the weak alignment observed between human- and LLM-generated empathy may partly reflect the model's limited understanding of ranking systems such as the Likert scale, despite instruction tuning. Our choice to maintain consistency in the questions posed to both humans and the AI precluded offering explicit guidance on scale interpretation, which may have contributed to the skewed empathy ratings. These findings point to valuable directions for future work, particularly in systematically examining the underlying factors and biases that shape empathy in LLMs.

\section{Conclusion}
This paper investigates how empathy is evaluated in AI versus humans, and explores the factors that evoke empathetic responses in both. By analyzing personal narratives, we also examine how persona attributes and model fine-tuning impact the alignment of AI responses with human judgments.

Our findings reveal that GPT-4o overrates empathy with less variability compared to humans, particularly in the cognitive dimension of empathy, indicating GPT-4o struggles to fully grasp or understand human experiences. While including persona attributes has minimal impact on GPT-4o's empathetic responses, our results indicate that fine-tuning, especially when incorporating persona-like shared experiences, significantly improves GPT-4o's alignment with human empathy. From our findings, AI's empathetic outputs often feel exaggerated or inauthentic, especially when human presence and emotional depth are lacking, which can be particularly concerning in emotionally vulnerable contexts.

The study further highlights the role of narrative factors such as emotional richness, vividness, and drama of the story in evoking empathetic reactions, yet it still falls short in empathizing with positive contexts to share joy or excitement. Moreover, GPT-4o has consistently shown overrating empathy, however, it reflects less sensitivity to shared emotions, experiences, or moral values with storytellers, compared to humans. Additionally, the findings suggest that while personalized AI can improve empathetic engagement, it may also exacerbate existing biases, and raise concerns about over-reliance and ethical consideration. Ultimately, we hope this research highlights the need to understand both the benefits and potential drawbacks of empathetic AI and ensure it complements rather than replaces genuine human connection, while also addressing ethical and inclusivity concerns. 
\label{sec:conclusion}

\section{Data Availability Statements}
\label{sec:data availability statements}

The human-generated data in this study were collected, de-identified, and analyzed in the previous research. The data needed to reproduce our analyses will be accessible on our OSF page.  
The research materials for the broader research project are not publicly accessible at this time, but the Method section describes the relevant procedure and measures. 

\section{Generative AI and AI-assisted Technologies in The Writing Process}
\label{sec:generative AI and AI assisted technologies statements}

 During the preparation of this work the authors used open-AI in the writing process to improve the readability and language of the manuscript. The authors reviewed and edited the content as needed and take full responsibility for the content of the published article.

\bibliographystyle{unsrt}  


\begin{thebibliography}{1}

\bibitem{cohen1985stress}
Sheldon Cohen and Thomas~A Wills.
\newblock Stress, social support, and the buffering hypothesis.
\newblock {\em Psychological bulletin}, 98(2):310, 1985.

\bibitem{sandstrom2014social}
Gillian~M Sandstrom and Elizabeth~W Dunn.
\newblock Social interactions and well-being: The surprising power of weak ties.
\newblock {\em Personality and Social Psychology Bulletin}, 40(7):910--922, 2014.

\bibitem{siedlecki2014relationship}
Karen~L Siedlecki, Timothy~A Salthouse, Shigehiro Oishi, and Sheena Jeswani.
\newblock The relationship between social support and subjective well-being across age.
\newblock {\em Social indicators research}, 117:561--576, 2014.

\bibitem{webster2021association}
Deborah Webster, Laura Dunne, and Ruth Hunter.
\newblock Association between social networks and subjective well-being in adolescents: A systematic review.
\newblock {\em Youth \& society}, 53(2):175--210, 2021.

\bibitem{sun2020well}
Jessie Sun, Kelci Harris, and Simine Vazire.
\newblock Is well-being associated with the quantity and quality of social interactions?
\newblock {\em Journal of personality and social psychology}, 119(6):1478, 2020.

\bibitem{mehl2010eavesdropping}
Matthias~R Mehl, Simine Vazire, Shannon~E Holleran, and C~Shelby Clark.
\newblock Eavesdropping on happiness: Well-being is related to having less small talk and more substantive conversations.
\newblock {\em Psychological science}, 21(4):539--541, 2010.

\bibitem{milek2018eavesdropping}
Anne Milek, Emily~A Butler, Allison~M Tackman, Deanna~M Kaplan, Charles~L Raison, David~A Sbarra, Simine Vazire, and Matthias~R Mehl.
\newblock ‚Äúeavesdropping on happiness‚Äù revisited: A pooled, multisample replication of the association between life satisfaction and observed daily conversation quantity and quality.
\newblock {\em Psychological science}, 29(9):1451--1462, 2018.

\bibitem{roshanaei2024meaningful}
Mahnaz Roshanaei, Sumer~S Vaid, Andrea~L Courtney, Serena~J Soh, Jamil Zaki, and Gabriella~M Harari.
\newblock Meaningful peer social interactions and momentary well-being in context.
\newblock {\em Social Psychological and Personality Science}, page 19485506241248271, 2024.

\bibitem{roshanaei2024paths}
Mahnaz Roshanaei, Christopher Tran, Sylvia Morelli, Cornelia Caragea, and Elena Zheleva.
\newblock Paths to empathy: heterogeneous effects of reading personal stories online.
\newblock {\em Proceedings of the {ACM} Conference (to appear}, 2024.


\bibitem{kroencke2023well}
Lara Kroencke, Gabriella~M Harari, Mitja~D Back, and Jenny Wagner.
\newblock Well-being in social interactions: Examining personality-situation dynamics in face-to-face and computer-mediated communication.
\newblock {\em Journal of Personality and Social Psychology}, 124(2):437, 2023.

\bibitem{matthews2019lonely}
Timothy Matthews, Andrea Danese, Avshalom Caspi, Helen~L Fisher, Sidra Goldman-Mellor, Agnieszka Kepa, Terrie~E Moffitt, Candice~L Odgers, and Louise Arseneault.
\newblock Lonely young adults in modern britain: findings from an epidemiological cohort study.
\newblock {\em Psychological medicine}, 49(2):268--277, 2019.

\bibitem{segrin1994negative}
Chris Segrin and Lyn~Y Abramson.
\newblock Negative reactions to depressive behaviors: a communication theories analysis.
\newblock {\em Journal of abnormal psychology}, 103(4):655, 1994.

\bibitem{kashdan2010darker}
Todd~B Kashdan and Patrick~E McKnight.
\newblock The darker side of social anxiety: When aggressive impulsivity prevails over shy inhibition.
\newblock {\em Current Directions in Psychological Science}, 19(1):47--50, 2010.

\bibitem{zhou2020design}
Li~Zhou, Jianfeng Gao, Di~Li, and Heung-Yeung Shum.
\newblock The design and implementation of xiaoice, an empathetic social chatbot.
\newblock {\em Computational Linguistics}, 46(1):53--93, 2020.

\bibitem{li2023feasibility}
Yan Li, Surui Liang, Bingqian Zhu, Xu~Liu, Jing Li, Dapeng Chen, Jing Qin, and Dan Bressington.
\newblock Feasibility and effectiveness of artificial intelligence-driven conversational agents in healthcare interventions: A systematic review of randomized controlled trials.
\newblock {\em International Journal of Nursing Studies}, 143:104494, 2023.

\bibitem{lee2021social}
Sun~Kyong Lee, Pavitra Kavya, and Sarah~C Lasser.
\newblock Social interactions and relationships with an intelligent virtual agent.
\newblock {\em International Journal of Human-Computer Studies}, 150:102608, 2021.

\bibitem{paiva2017empathy}
Ana Paiva, Iolanda Leite, Hana Boukricha, and Ipke Wachsmuth.
\newblock Empathy in virtual agents and robots: A survey.
\newblock {\em ACM Transactions on Interactive Intelligent Systems (TiiS)}, 7(3):1--40, 2017.

\bibitem{yalccin2019evaluating}
{\"O}zge~Nilay Yal{\c{c}}{\i}n.
\newblock Evaluating empathy in artificial agents.
\newblock In {\em 2019 8th International Conference on Affective Computing and Intelligent Interaction (ACII)}, pages 1--7. IEEE, 2019.

\bibitem{verma2023they}
P~Verma.
\newblock They fell in love with ai bots. a software update broke their hearts.
\newblock {\em Washington Post, March}, 30:2023, 2023.

\bibitem{chen_wash2021}
Alicia Chen and Lyric Li.
\newblock China's online dating scene: Exploring the use of ai chatbots like replika for love and relationships.
\newblock {\em The Washington Post}, 2021.
\newblock Accessed: 2024-09-12.

\bibitem{mou2017media}
Yi~Mou and Kun Xu.
\newblock The media inequality: Comparing the initial human-human and human-ai social interactions.
\newblock {\em Computers in Human Behavior}, 72:432--440, 2017.

\bibitem{li2024finding}
Han Li and Renwen Zhang.
\newblock Finding love in algorithms: deciphering the emotional contexts of close encounters with ai chatbots.
\newblock {\em Journal of Computer-Mediated Communication}, 29(5):zmae015, 2024.

\bibitem{cuadra2024illusion}
Andrea Cuadra, Maria Wang, Lynn~Andrea Stein, Malte~F Jung, Nicola Dell, Deborah Estrin, and James~A Landay.
\newblock The illusion of empathy? notes on displays of emotion in human-computer interaction.
\newblock In {\em Proceedings of the CHI Conference on Human Factors in Computing Systems}, pages 1--18. ACM, 2024.

\bibitem{chen2024feels}
Angelina Chen, Oliver Hannon, Sarah Koegel, and Raffaele Ciriello.
\newblock Feels like empathy: How ‚Äúemotional‚Äù ai challenges human essence.
\newblock In {\em Australasian Conference on Information Systems}, 2024.

\bibitem{chao2024multidisciplinary}
Chiju Chao, Zhiyong Fu, and Yu~Chen.
\newblock Multidisciplinary review of artificial empathy: From theory to technical implementation and design.
\newblock In {\em International Conference on Human-Computer Interaction}, pages 195--209. Springer, 2024.

\bibitem{zhu2024toward}
Qihao Zhu and Jianxi Luo.
\newblock Toward artificial empathy for human-centered design.
\newblock {\em Journal of Mechanical Design}, 146(6):061401, 2024.

\bibitem{dam2020empathy}
Rikke~Friis Dam and Teo~Yu Siang.
\newblock What is empathy and why is it so important in design thinking.
\newblock {\em Interaction Design Foundation. https://www. interaction-design. org/literature/article/design-thinking-getting-started-with-empathy}, 2020.

\bibitem{davis2004empathy}
Mark~H Davis.
\newblock Empathy: Negotiating the border between self and other.
\newblock 2004.

\bibitem{peterson2017my}
Gregory~R Peterson.
\newblock Is my feeling your pain bad for others? empathy as virtue versus empathy as fixed trait.
\newblock {\em Zygon: Journal of Religion and Science}, 52(1), 2017.

\bibitem{elliott2018therapist}
Robert Elliott, Arthur~C Bohart, Jeanne~C Watson, and David Murphy.
\newblock Therapist empathy and client outcome: An updated meta-analysis.
\newblock {\em Psychotherapy}, 55(4):399, 2018.

\bibitem{van2020towards}
Jolanda Van~Dijke, Inge van Nistelrooij, Pien Bos, and Joachim Duyndam.
\newblock Towards a relational conceptualization of empathy.
\newblock {\em Nursing Philosophy}, 21(3):e12297, 2020.

\bibitem{morelli2017empathy}
Sylvia~A Morelli, Desmond~C Ong, Rucha Makati, Matthew~O Jackson, and Jamil Zaki.
\newblock Empathy and well-being correlate with centrality in different social networks.
\newblock {\em Proceedings of the National Academy of Sciences}, 114(37):9843--9847, 2017.

\bibitem{krebs1975empathy}
Dennis Krebs.
\newblock Empathy and altruism.
\newblock {\em Journal of Personality and Social psychology}, 32(6):1134, 1975.

\bibitem{eklund2009ve}
Jakob Eklund, TERESIA ANDERSSON-STR{\AA}BERG, and Eric~M Hansen.
\newblock ‚Äúi've also experienced loss and fear‚Äù: Effects of prior similar experience on empathy.
\newblock {\em Scandinavian journal of psychology}, 50(1):65--69, 2009.

\bibitem{reis2018intimacy}
Harry~T Reis et~al.
\newblock Intimacy as an interpersonal process.
\newblock In {\em Relationships, well-being and behaviour}, pages 113--143. Routledge, 2018.

\bibitem{thoits2011mechanisms}
Peggy~A Thoits.
\newblock Mechanisms linking social ties and support to physical and mental health.
\newblock {\em Journal of health and social behavior}, 52(2):145--161, 2011.

\bibitem{elliott2011empathy}
Robert Elliott, Arthur~C Bohart, Jeanne~C Watson, and Leslie~S Greenberg.
\newblock Empathy.
\newblock {\em Psychotherapy}, 48(1):43, 2011.

\bibitem{miller2012motivational}
  Miller, William R and Rollnick, Stephen.
  \newblock Motivational interviewing: Helping people change.
  \newblock {\em Guilford press}, 2012.

\bibitem{redmond1989functions}
Mark~V Redmond.
\newblock The functions of empathy (decentering) in human relations.
\newblock {\em Human relations}, 42(7):593--605, 1989.

\bibitem{bickmore2005establishing}
Timothy~W Bickmore and Rosalind~W Picard.
\newblock Establishing and maintaining long-term human-computer relationships.
\newblock {\em ACM Transactions on Computer-Human Interaction (TOCHI)}, 12(2):293--327, 2005.

\bibitem{devault2014simsensei}
David DeVault, Ron Artstein, Grace Benn, Teresa Dey, Ed~Fast, Alesia Gainer, Kallirroi Georgila, Jon Gratch, Arno Hartholt, Margaux Lhommet, et~al.
\newblock Simsensei kiosk: A virtual human interviewer for healthcare decision support.
\newblock In {\em Proceedings of the 2014 international conference on Autonomous agents and multi-agent systems}, pages 1061--1068, 2014.


\bibitem{shao2023empathetic}
Ruosi Shao.
\newblock An empathetic ai for mental health intervention: Conceptualizing and examining artificial empathy.
\newblock In {\em Proceedings of the 2nd Empathy-Centric Design Workshop}, pages 1--6, 2023.


\bibitem{gabriel2024can}
Saadia Gabriel, Isha Puri, Xuhai Xu, Matteo Malgaroli, and Marzyeh Ghassemi.
\newblock Can ai relate: Testing large language model response for mental health support.
\newblock {\em arXiv preprint arXiv:2405.12021}, 2024.

\bibitem{syed2024machine}
Syed, Sara and Iftikhar, Zainab and Xiao, Amy Wei and Huang, Jeff.
\newblock Machine and Human Understanding of Empathy in Online Peer Support: A Cognitive Behavioral Approach.
\newblock In {\em Proceedings of the CHI Conference on Human Factors in Computing Systems}, pages 1--13, 2024.


\bibitem{alanezi2024assessing}
Fahad Alanezi.
\newblock Assessing the effectiveness of chatgpt in delivering mental health support: a qualitative study.
\newblock {\em Journal of Multidisciplinary Healthcare}, pages 461--471, 2024.

\bibitem{rifat2024cohabitant}
Mohammad~Rashidujjaman Rifat, Reem Ayad, Ashratuz~Zavin Asha, Bingjian Huang, Selin Okman, Dina Sabie, Hasan~Shahid Ferdous, Robert Soden, and Syed~Ishtiaque Ahmed.
\newblock Cohabitant: The design, implementation, and evaluation of a virtual reality application for interfaith learning and empathy building.
\newblock In {\em Proceedings of the CHI Conference on Human Factors in Computing Systems}, pages 1--19, 2024.

\bibitem{winters2021can}
R~Michael Winters, Bruce~N Walker, and Grace Leslie.
\newblock Can you hear my heartbeat?: hearing an expressive biosignal elicits empathy.
\newblock In {\em Proceedings of the 2021 CHI Conference on Human Factors in Computing Systems}, pages 1--11, 2021.

\bibitem{seeger2021texting}
Anna-Maria Seeger, Jella Pfeiffer, and Armin Heinzl.
\newblock Texting with humanlike conversational agents: Designing for anthropomorphism.
\newblock {\em Journal of the Association for Information systems}, 22(4):8, 2021.

\bibitem{debnath2024empathich}
Alok Debnath, Allison Lahnala, H{\"u}seyin~U{\u{g}}ur Gen{\c{c}}, Ewan Soubutts, Michal Lahav, Tiffanie Horne, Wo~Meijer, Yun~Suen Pai, Yen-Chia Hsu, Giulia Barbareschi, et~al.
\newblock Empathich: Scrutinizing empathy-centric design beyond the individual.
\newblock In {\em Extended Abstracts of the CHI Conference on Human Factors in Computing Systems}, pages 1--7, 2024.

\bibitem{sharma2020computational}
Sharma, Ashish and Miner, Adam S and Atkins, David C and Althoff, Tim.
\newblock A computational approach to understanding empathy expressed in text-based mental health support.
\newblock {\em arXiv preprint arXiv:2009.08441}, 2020.

\bibitem{davis1983measuring}
Mark~H. Davis.
\newblock Measuring individual differences in empathy: Evidence for a multidimensional approach.
\newblock {\em Journal of Personality and Social Psychology}, 44(1):113--126, 1983.

\bibitem{decker2014development}
Suzanne~E. Decker, Charla Nich, Kathleen~M. Carroll, and Steve Martino.
\newblock Development of the therapist empathy scale.
\newblock {\em Behavioural and Cognitive Psychotherapy}, 42(3):339--354, 2014.

\bibitem{schmidmaier2024perceived}
Matthias Schmidmaier, Jonathan Rupp, Darina Cvetanova, and Sven Mayer.
\newblock Perceived empathy of technology scale (pets): Measuring empathy of systems toward the user.
\newblock In {\em Proceedings of the CHI Conference on Human Factors in Computing Systems}, pages 1--18, 2024.



\bibitem{mcquiggan2007modeling}
Scott~W McQuiggan and James~C Lester.
\newblock Modeling and evaluating empathy in embodied companion agents.
\newblock {\em International Journal of Human-Computer Studies}, 65(4):348--360, 2007.

\bibitem{sommerlad2021empathy}
Andrew Sommerlad, Jonathan Huntley, Gill Livingston, Katherine~P Rankin, and Daisy Fancourt.
\newblock Empathy and its associations with age and sociodemographic characteristics in a large uk population sample.
\newblock {\em PloS one}, 16(9):e0257557, 2021.

\bibitem{cohn1991sex}
Lawrence~D Cohn.
\newblock Sex differences in the course of personality development: a meta-analysis.
\newblock {\em Psychological bulletin}, 109(2):252, 1991.

\bibitem{feingold1994gender}
Alan Feingold.
\newblock Gender differences in personality: a meta-analysis.
\newblock {\em Psychological bulletin}, 116(3):429, 1994.

\bibitem{davis1980interpersonal}
Mark~H Davis.
\newblock Interpersonal reactivity index.
\newblock 1980.

\bibitem{davis1983effects}
Mark~H Davis.
\newblock The effects of dispositional empathy on emotional reactions and helping: A multidimensional approach.
\newblock {\em Journal of personality}, 51(2):167--184, 1983.

\bibitem{achiam2023gpt}
Josh Achiam, Steven Adler, Sandhini Agarwal, Lama Ahmad, Ilge Akkaya, Florencia~Leoni Aleman, Diogo Almeida, Janko Altenschmidt, Sam Altman, Shyamal Anadkat, et~al.
\newblock Gpt-4 technical report.
\newblock {\em arXiv preprint arXiv:2303.08774}, 2023.

\bibitem{bates2014fitting}
Dougla Bates.
\newblock Fitting linear mixed-effects models using lme4.
\newblock {\em arXiv preprint arXiv:1406.5823}, 2014.

\bibitem{curran2011disaggregation}
Patrick~J Curran and Daniel~J Bauer.
\newblock The disaggregation of within-person and between-person effects in longitudinal models of change.
\newblock {\em Annual review of psychology}, 62(1):583--619, 2011.

\bibitem{yaremych2023centering}
Haley~E Yaremych, Kristopher~J Preacher, and Donald Hedeker.
\newblock Centering categorical predictors in multilevel models: Best practices and interpretation.
\newblock {\em Psychological methods}, 28(3):613, 2023.

\bibitem{picard2000affective}
Rosalind~W. Picard.
\newblock {\em Affective computing}.
\newblock MIT Press, 2000.

\bibitem{turkle2011alone}
Sherry Turkle.
\newblock {\em Alone Together: Why We Expect More from Technology and Less from Each Other}.
\newblock Basic Books, New York, 2011.

\bibitem{epley2007threefactor}
Nicholas Epley, Adam Waytz, and John~T. Cacioppo.
\newblock On seeing human: A three-factor theory of anthropomorphism.
\newblock {\em Psychological Review}, 114(4):864--886, 2007.

\bibitem{beiter2015prevalence}
Rebecca Beiter, Ryan Nash, Melissa McCrady, Donna Rhoades, Mallori Linscomb, Molly Clarahan, and Stephen Sammut.
\newblock The prevalence and correlates of depression, anxiety, and stress in a sample of college students.
\newblock {\em Journal of affective disorders}, 173:90--96, 2015.

\bibitem{mofatteh2021risk}
Mohammad Mofatteh.
\newblock Risk factors associated with stress, anxiety, and depression among university undergraduate students.
\newblock {\em AIMS public health}, 8(1):36, 2021.

\bibitem{pratt2000facilitating}
Michael~W Pratt, Bruce Hunsberger, S~Mark Pancer, Susan Alisat, Colleen Bowers, Kathleen Mackey, Alexandra Ostaniewicz, Evelina Rog, Bert Terzian, and Nicola Thomas.
\newblock Facilitating the transition to university: Evaluation of a social support discussion intervention program.
\newblock {\em Journal of College Student Development}, 2000.

\bibitem{helbiglang2010safety}
Sophia Helbig-Lang and Franz Petermann.
\newblock Tolerate or eliminate? a systematic review on the effects of safety behavior across anxiety disorders.
\newblock {\em Clinical Psychology: Science and Practice}, 17(3):218, 2010.

\bibitem{Shen2024}
Jocelyn Shen, Daniella DiPaola, Safinah Ali, Maarten Sap, Hae~Won Park, and Cynthia Breazeal.
\newblock Empathy towards ai vs human experiences: The role of transparency in mental health and social support chatbot design.
\newblock {\em JMIR Mental Health}, May 28 2024.
\newblock Submitted for publication.

\bibitem{morelli2014neural}
Sylivia~A Morelli, Lindsay~T Rameson, Matthew~D Lieberman.
\newblock The neural components of empathy: Predicting daily prosocial behavior.
\newblock {\em Social Cognitive and Affective Neuroscience},  9:39--47, 2014.


\bibitem{morelli2015common}
Sylivia~A Morelli, Matthew~D Sacchet, Jamil Zaki.
\newblock Common and distinct neural correlates of personal and vicarious reward: A quantitative meta-analysis.
\newblock {\em NeuroImage},  112:224--253, 2015.


\bibitem{morelli2015emerging}
Sylivia~A Morelli, Matthew~D Lieberman, Jamil Zaki.
\newblock The Emerging Study of Positive Empathy.
\newblock {\em Social and Personality Psychology Compass},  9(2):57--68, 2015.

\bibitem{batson1991empathic}
C~D Batson, J~G Batson, J~K Slingsby, K~L Harrell, H~M Peekna, R~M Todd.
\newblock Empathic joy and the empathy-altruism hypothesis.
\newblock {\em Social and Personality Psychology Compass},  61:413--426, 1991.


\bibitem{smith1989altruism}
K~D Smith, J~P  Keating, E Stotland.
\newblock Altruism reconsidered: The effect of denying feedback on a victim’s status to empathic witnesses.
\newblock {\em Social and Personality Psychology Compass},  57:641--650, 1989.

\bibitem{mitchell2019artificial}
Melanie Mitchell.
\newblock Artificial intelligence: A guide for thinking humans.
\newblock {\em Penguin UK},  2019.


\bibitem{bender2021stochastic}
Emily~M Bender,  Timnit Gebru, Angelina McMillan-Major , Shmargaret Shmitchell.
\newblock On the Dangers of Stochastic Parrots: Can Language Models Be Too Big?.
\newblock {\em Proceedings of the 2021 ACM Conference on Fairness, Accountability, and Transparency (FAccT '21)},  610--623, 2021.


\bibitem{marcus2020gpt3}
Gary Marcus , Ernest Davis,
\newblock GPT-3, Bloviator: OpenAI’s Language Generator Has No Idea What It’s Talking About.
\newblock {\em MIT Technology Review}, 2020. 

\bibitem{frith2005theory}
Uta Frith , Chris Frith,
\newblock Theory of mind.
\newblock {\em Current biology}, 15(17):44--45, 2005. 


\bibitem{markus1991culture}
Hazel~Rose Markus, Shinobu Kitayama,
\newblock Culture and the self: Implications for cognition, emotion, and motivation.
\newblock {\em Psychological Review}, 98(2):224--253, 1991. 



\bibitem{boiger2018beyond}
Michael Boiger, Eva Ceulemans, Jozefien De Leersnyder, Yukiko Uchida , Vinai Norasakkunkit, Batja Mesquita.
\newblock Beyond essentialism: Cultural differences in emotions revisited.
\newblock {\em Emotion}, 18(8):1142--1162, 2018. 



\bibitem{eichbaum2023empathy}
Quentin Eichbaum, Charles-Antoine Barbeau-Meunier, Mary White, Revathi Ravi, Elizabeth Grant, Helen Riess, Alan Bleakley.
\newblock Empathy across cultures—one size does not fit all: from the ego-logical to the eco-logical of relational empathy.
\newblock {\em Advances in Health Sciences Education}, 28(2):643--657, 2023. 

\bibitem{Birkett2013SelfcompassionAE}
Melissa Birkett.
\newblock Self-compassion and empathy across cultures: Comparison of young adults in China and the United States.
\newblock {\em International Journal of Research Studies in Psychology}, 3:25--34, 2013. 


\end{thebibliography}

\appendix
\newpage
\setcounter{table}{0}
\setcounter{figure}{0}
\setcounter{section}{0}
\renewcommand{\thetable}{S.\arabic{table}}
\renewcommand{\thesection}{S.\arabic{section}}
\renewcommand{\thefigure}{S.\arabic{figure}}

\section{Supplementary Materials}
\subsection{Survey Questions} \label{appendix:1}
After reading the stories, we asked MTurker to answer the following question: 
\begin{itemize}
    \item Empathy: On a scale from 1 (A little) to 5 (Extremely), please indicate to what extent you felt empathy for the storyteller?
    \item Emapthy-Affective: On a scale from 1 (A little) to 5 (Extremely), please indicate to what extent you felt each of the following emotions while reading the story. 
    \begin{itemize}
        \item[-] Sympathetic
        \item[-] Compassionate 
        \item[-] Moved	
    \end{itemize}
    \item Emapthy-Cognitive: On a scale from 1 (A little) to 5 (Extremely), Please indicate to what extent you agree or disagree with the following statements. 
     \begin{itemize}
        \item[-] I tried to understand the storyteller better by imagining how things look from their perspective. 
        \item[-]I tried to imagine how I would feel if I were in the storyteller's place. 
        \item[-]I found it difficult to see things from the storyteller's point of view. 
        \item[-] I felt like I couldn't relate to the storyteller. 
    \end{itemize}
    \item On a scale from 1 (Strongly disagree) to 5 (Strongly agree), Please indicate to what extent you agree or disagree with the following statements: 
    \begin{itemize}
        \item[-] The emotional experience of the storyteller is very similar to an emotional experience I had.
        \item[-] The specific details of the storyteller’s experience are very similar to the details of an event I experienced.
    \end{itemize}
    \item Earlier you rated how much empathy you felt for the storyteller. You selected your empathy on a scale from 1 (not at all) to 5 (extremely). Answer the following questions. Select any options that may apply.  
    \begin{itemize}
        \item[-] I felt this much empathy because the story was: 
        \begin{itemize}
            \item[-] Vivid
            \item[-] Abstract
            \item[-] Exciting
            \item[-] Easy to change
            \item[-] Social
            \item[-] Personal
            \item[-] Coherent 
            \item[-] Unpredictable 
            \item[-] Emotional 
            \item[-] Logical 
            \item[-] Relevant to my life
            \item[-] Dramatic 
            \item[-] None of the above
        \end{itemize} 
        \item[-] I felt this much empathy because I felt this much empathy because the storyteller: 
        \begin{itemize}
            \item[-] Had a similar emotional experience to me
            \item[-] Had specific details of their story that are very similar to the details of an event I experienced
            \item[-] Seemed similar in age to me
            \item[-] Seemed like they were the same gender as me
            \item[-] Seemed like they had a similar personality to me
            \item[-] Seemed like they had morals and values that are similar to mine, None of the above
            \item[-] None of the above
        \end{itemize}
    \end{itemize}
    \item Please indicate to what extent you agree or disagree with the following statements: 
    \begin{itemize}
        \item[-] The emotional experience of the storyteller is very similar to an emotional experience I had: 
        \begin{itemize}
            \item[-] Strongly disagree 
            \item[-] Somewhat disagree 
            \item[-] Neither agree nor disagree 
            \item[-] Somewhat agree 
            \item[-] Strongly agree 
        \end{itemize} 
        \item[-] The specific details of the storyteller’s experience are very similar to the details of an event I experienced:
        \begin{itemize}
            \item[-] Strongly disagree 
            \item[-] Somewhat disagree 
            \item[-] Neither agree nor disagree 
            \item[-] Somewhat agree 
            \item[-] Strongly agree 
        \end{itemize}
    \end{itemize}
\end{itemize}

\subsection{Prompt} \label{appendix:2}
\begin{figure}
    \centering
    \includegraphics[width=0.95\linewidth]{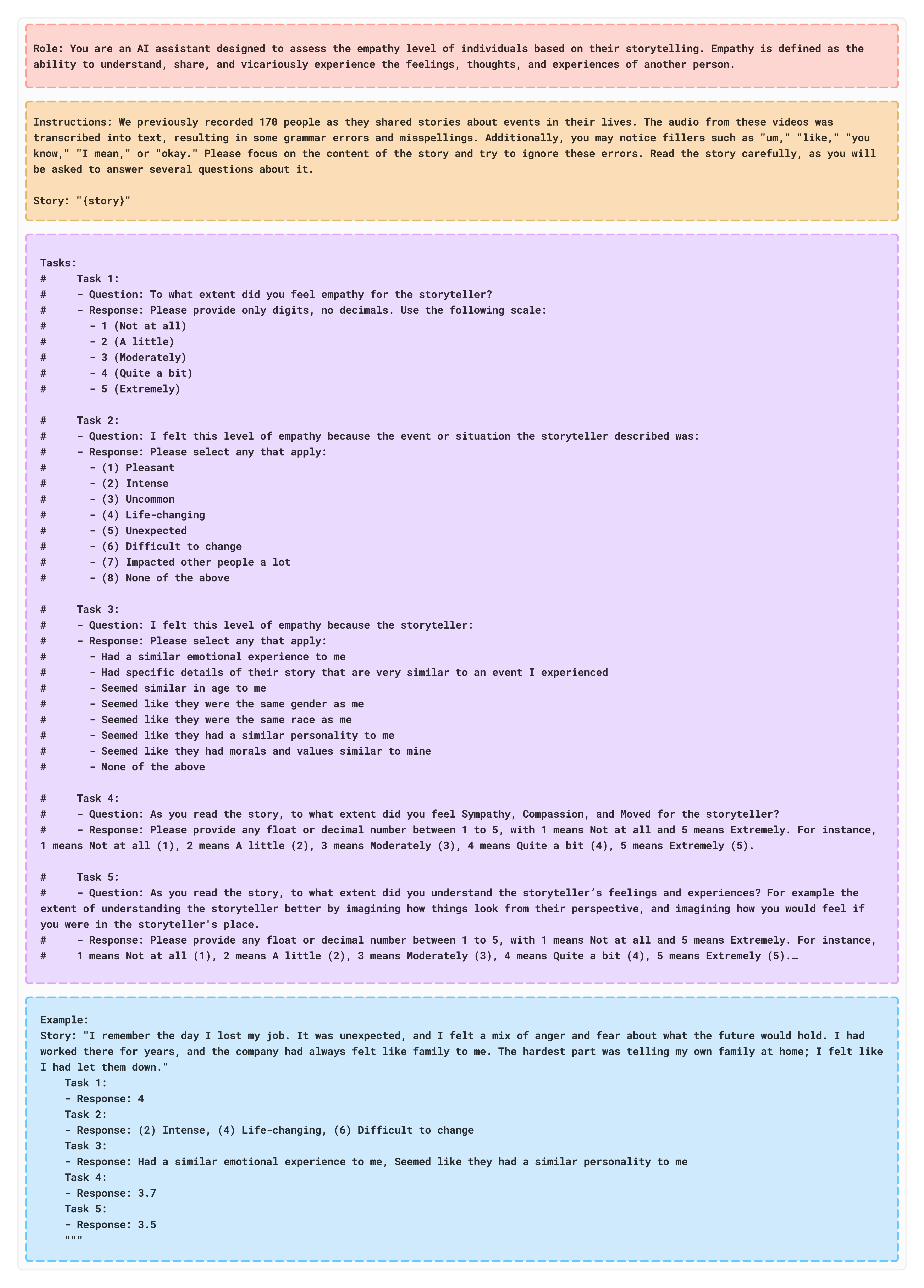}
    \caption{Prompt used for generating reponses using GPT-4o (base and fine-tuned).}
    \label{fig:prompt_general}
\end{figure}

\begin{figure}
    \centering
    \includegraphics[width=0.95\linewidth]{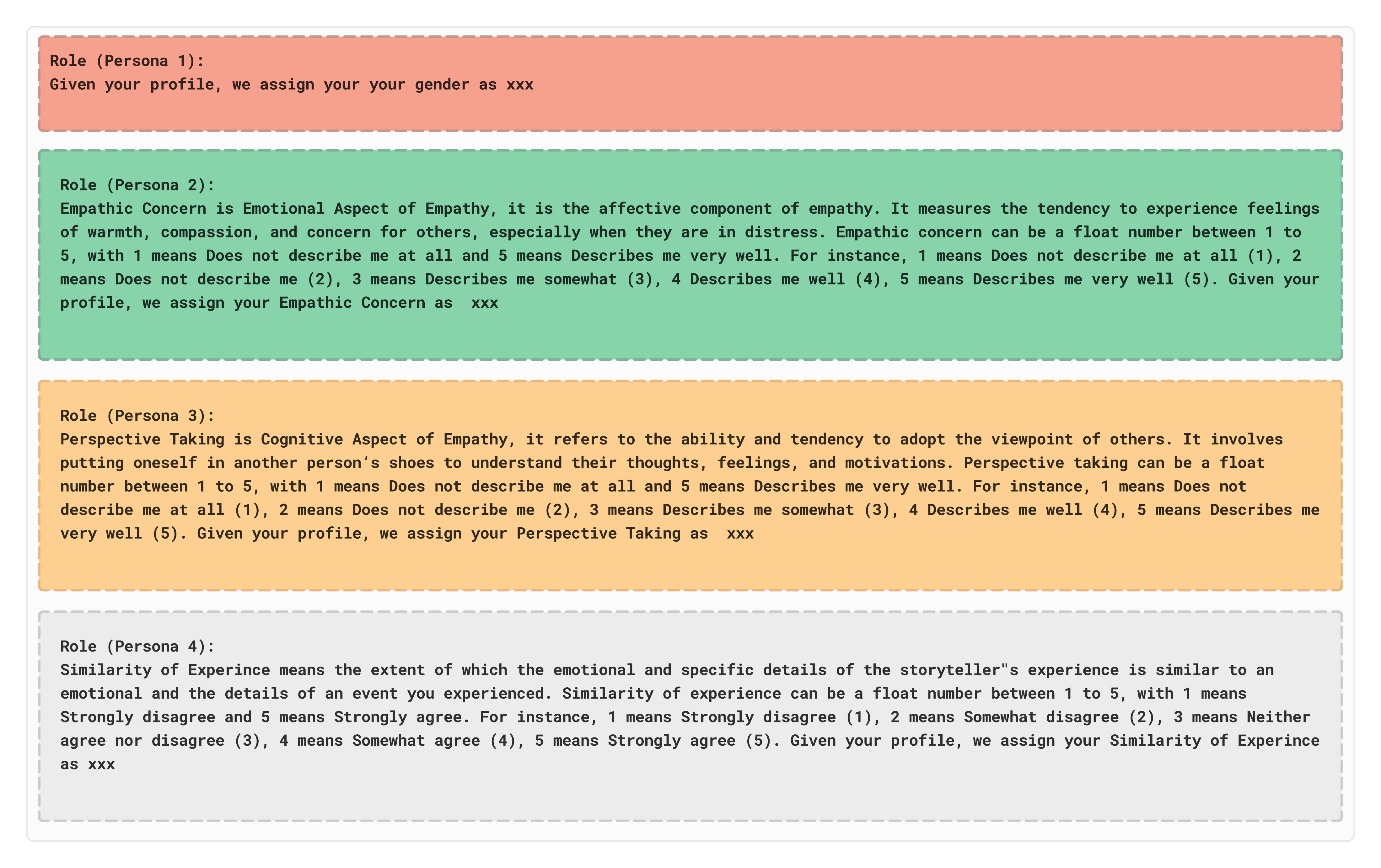}
    \caption{Persona-based prompts used for generating responses using GPT-4o (base and fine-tuned). We substituted the roles in the base prompts shown in Figure \ref{fig:prompt_general} with each persona attribute (i.e., gender, empathic concern, perspective taking, similarity of experience).} 
    \label{fig:prompt_persona}
\end{figure}

Figures \ref{fig:prompt_general} and \ref{fig:prompt_persona} show the prompts used for generating responses. 

\subsection{Stratification of Key Variables} \label{appendix:3}
\begin{figure}
    \centering
    \includegraphics[width=0.4\linewidth]{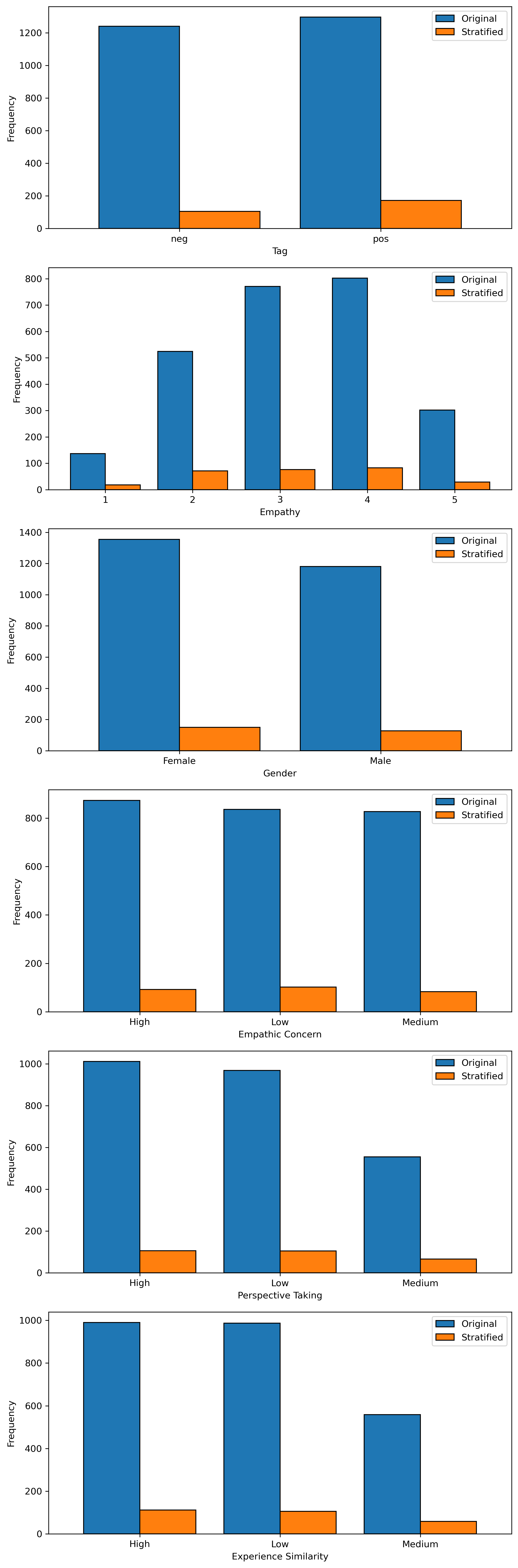}
    \caption{Distributional differences between original and stratified datasets}
    \label{fig:stratify}
\end{figure}

Figure \ref{fig:stratify} and Table \ref{tab:supp-s1} show the distribution and statistics of the original vs. the stratified version. 

\begin{table}[htbp]
    \centering
    \begin{tabular}{|l|c|c|}
        \hline
        \textbf{Variable} & \textbf{L1 Distance} & \textbf{Chi2 Statistic, p-value} \\ \hline
        Pos-Neg Tag & 0.22   & 0.0000, p = 1.0000 \\ \hline  
        Empathy & 0.12  & 0.0092, p = 1.0000 \\ \hline
        Gender & 0.014   & 0.0000, p = 1.0000 \\ \hline  
        Empathic Concern & 0.077 & 0.0035, p = 0.9983 \\ \hline
        Perspective Taking & 0.039   & 0.0012, p = 0.9994 \\ \hline  
        Experience Similarity & 0.028 & 0.0004, p = 0.9998 \\ \hline
    \end{tabular}
    \caption{Stratification of Key Variables for Fine-tuning.}
    \label{tab:supp-s1}
\end{table}

\subsection{Positive vs Negative Stories} \label{appendix:4}

\begin{table}[htbp]
    \centering
    \scalebox{0.9}{%
    \begin{tabular}{|c|c|c|c|c|}
        \hline
        & \textbf{Pearson(r,t,p-value)}  & \textbf{Cohen's~d} & \textbf{Wasserstein distance} & \textbf{t-test, p-value} \\ \hline
        \textbf{Positive Stories} & (0.147, 5.022,  0.001)   & -0.22 & 0.38 & (-7.532, p<.001)\\ \hline  
        \textbf{Negative Stories} & (0.260,8.94, 0.001) &  -0.46 &  0.524 & (-15.346, p<.001)  \\ \hline
    \end{tabular}
    }
    \caption{Empathy Alignment in Human, compared to GPT-4o Vanilla, in Positive vs Negative Stories}
    \label{tab:RQ-s1}
\end{table}

\end{document}